\documentclass[11pt,a4paper]{article}

\usepackage{jheppub}
\usepackage{graphicx}
\usepackage{dcolumn}
\usepackage{bm}
\usepackage{amsmath}
\usepackage{braket}
\usepackage{slashed}    
\usepackage{epstopdf}
\usepackage{placeins}
\usepackage{multirow}
\usepackage{makecell}
\usepackage[shortlabels]{enumitem}
\usepackage{cleveref}
\usepackage{xcolor}

\usepackage[font=small, labelfont=bf, textfont={small,it}]{caption}
\newcommand{\beq}{\begin{eqnarray}}
\newcommand{\eeq}{\end{eqnarray}}
\newcommand{\be}{\begin{equation}}
\newcommand{\ee}{\end{equation}}
\newcommand{\beqnn}{\begin{eqnarray*}}
\newcommand{\eeqnn}{\end{eqnarray*}}
\newcommand{\Tr}{\ensuremath{\mathrm{Tr}}}
\newcommand{\SU}{\mathrm{SU}}
\newcommand{\YM}{\mathrm{YM}}
\newcommand{\clov}{\mathrm{clov}}

\newcommand{\MS}{\overline{\mathrm{MS}}}
\newcommand{\ov}{\mathrm{ov}}
\newcommand{\TGF}{\mathrm{TGF}}
\newcommand{\PTBC}{\mathrm{PTBC}}

\newcommand{\s}{\mathrm{s}}
\newcommand{\tl}{\tilde l}
\newcommand{\tL}{\tilde L}
\newcommand{\had}{\mathrm{had}}
\newcommand{\pt}{\mathrm{pt}}
\newcommand{\noproj}{\mathrm{noproj}}
\newcommand{\tg}{\mathrm{tg}}

\begin{document}

\title{The $\mathrm{SU}(3)$ twisted gradient flow strong coupling without topology freezing}

\author[a]{Claudio Bonanno,}
\emailAdd{claudio.bonanno@csic.es}

\author[b]{Jorge Luis Dasilva Gol\'an,}
\emailAdd{jgolandas@bnl.gov}

\author[c]{Massimo D'Elia,}
\emailAdd{massimo.delia@unipi.it}

\author[a]{Margarita Garc\'ia P\'erez,}
\emailAdd{margarita.garcia@csic.es}

\author[c]{Andrea Giorgieri}
\emailAdd{andrea.giorgieri@phd.unipi.it}

\affiliation[a]{Instituto de F\'isica T\'eorica UAM-CSIC, Calle Nicol\'as Cabrera 13-15,\\Universidad Aut\'onoma de Madrid, Cantoblanco, E-28049 Madrid, Spain}

\affiliation[b]{Physics Deparment, Brookhaven National Laboratory, Upton, New York 11973, USA}

\affiliation[c]{Dipartimento di Fisica dell'Universit\`a di Pisa \& \\ INFN Sezione di Pisa, Largo Pontecorvo 3, I-56127 Pisa, Italy}

\abstract{We investigate the role of topology on the lattice determination of the $\mathrm{SU}(3)$ strong coupling renormalized via gradient flow. To deal with the topological freezing of standard local algorithms, the definition of the coupling is usually projected onto the zero topological sector. However, it is not obvious that this definition is not biased by the loss of ergodicity. We instead avoid the topological freezing using a novel algorithm, the Parallel Tempering on Boundary Conditions. The comparison with a standard algorithm shows that, even in the case where the latter is severely frozen, one obtains the same projected coupling. Moreover, we show that the two definitions of the coupling, projected and non-projected, lead to the same flow of the renormalization scale. Our results imply that projecting the coupling does not affect the determination of the dynamically-generated scale of the theory $\Lambda$, as obtained through the step-scaling method.}

\keywords{Lattice Gauge Theories, Renormalization}

\maketitle
\flushbottom

\section{Introduction}\label{sec:intro}

It is widely acknowledged that the Standard Model of Particle Physics cannot provide satisfactory explanations for various experimental observations. These include neutrino masses, Dark Matter, and strong-CP conservation, among others. As a result, the search for Physics beyond the Standard Model has been the focus of significant theoretical and experimental efforts in recent decades. Research in this field has led to the need for more precise and refined theoretical predictions of experimentally measurable quantities within the framework of the Standard Model itself.

In this respect, it has been emphasized that reducing the theoretical uncertainty on the strong coupling constant $\alpha_{\rm strong} = g^2/(4\pi)$ will be crucial in the study of several physical processes in the near future, see, e.g., Ref.~\cite{DallaBrida:2020pag} for a recent review. In the last two decades, the Lattice Community has spent a huge effort to improve the precision of the determination of the strong coupling~\cite{Maltman:2008bx,PACS-CS:2009zxm,McNeile:2010ji,Chakraborty:2014aca,Bruno:2017gxd,Cali:2020hrj,Bazavov:2019qoo,Ayala:2020odx}. As a result, the averaged lattice estimation~\cite{FlavourLatticeAveragingGroupFLAG:2021npn} is now among the most accurate determinations entering the world-average reported in the PDG~\cite{Workman:2022ynf}.

From the lattice perspective, determining the strong coupling constant practically amounts to the calculation of the dynamically generated scale $\Lambda_{\rm QCD}$. Thanks to the so-called \emph{decoupling method}~\cite{DallaBrida:2019mqg,DallaBrida:2020pag,DelDebbio:2021ryq}, this can be in turn traced back to the computation of the confinement scale $\Lambda_{\YM}$ of the pure-gauge theory, i.e., the pure $\SU(3)$ gluodynamics with no dynamical quarks. This quantity has been the target of several lattice calculations in the last 15 years~\cite{Brambilla:2010pp,Asakawa:2015vta,Kitazawa:2016dsl,Ishikawa:2017xam,Husung:2017qjz,DallaBrida:2019wur,Nada:2020jay,Husung:2020pxg,Bribian:2021cmg,Hasenfratz:2023bok}.

At first glance, the determination of $\Lambda_{\YM}$ seems a trivial task compared to the calculation of $\Lambda_{\rm QCD}$, given that pure-Yang--Mills simulations are less computationally demanding than full QCD ones involving dynamical fermions. However, the accurate determination of this quantity presents its own numerical challenges, as it requires keeping several sub-percent uncertainties under control.
This paper addresses one potential source of systematic errors: the strong correlation between the renormalized coupling and the topological modes of Yang-Mills theories.

To understand the reason behind this correlation and why it can be an issue, consider the following. A powerful and accurate technique to determine the renormalized strong coupling from lattice simulations is to define it from the action density after the gauge fields have been evolved under the gradient flow~\cite{Narayanan:2006rf,Lohmayer:2011si,Luscher:2009eq}. After the flow, the action density becomes highly correlated with the topological background of the underlying gauge field~\cite{Fritzsch:2013yxa}. The topological charge $Q$, in turn, suffers for very large auto-correlation times if the lattice spacing is fine~\cite{Alles:1996vn,DelDebbio:2004xh,Schaefer:2010hu}. This computational problem, known as \emph{topological freezing}, is due to the loss of ergodicity of standard local algorithms close to the continuum limit. In practice, it prevents sampling correctly the topological charge distribution in affordable Monte Carlo simulations. As a matter of fact, when the lattice spacing is sufficiently fine, very few to no fluctuations of $Q$ get sampled during typical runs. However, exploring fine lattices is necessary to pin down the systematic error related to the continuum limit extrapolation. So, given the strong correlation between topological charge and flowed action density, topological freezing can introduce a bias in the determinations of the renormalized coupling and, therefore, of the $\Lambda$-parameter.

Concerning the lattice determination of $\Lambda$, the freezing problem is usually circumvented by defining the coupling through a projection onto the $Q=0$ topological charge sector~\cite{Fritzsch:2013yxa}.\footnote{For another possible strategy to avoid topological freezing without $Q=0$ projection based on the combined use of Schr\"odinger Functional and open boundary conditions see Ref.~\cite{Luscher:2014kea}.} This choice can be seen just as a particular renormalization scheme. As such, after the proper matching with a customary scheme such as the $\MS$, the projection should not introduce any systematic in the determination of $\Lambda$. However, this approach has been the object of debate within the Lattice Community, and in particular two main issues have been raised.

First, when topological freezing is present, non-trivial fluctuations (e.g., instanton--anti-instanton pairs) in each fixed topological sector could get sampled incorrectly.\footnote{For a first investigation of this issue in the $2d$ $\mathrm{U}(1)$ gauge theory see Ref.~\cite{Albandea:2021lvl}.} If this were the case, topological freezing would introduce an ergodicity problem in the sampling of the $Q=0$ sector, possibly biasing the strong coupling determination obtained via projection. Second, the $\Lambda$-parameter is customarily computed via the \emph{step-scaling} technique, which consists in flowing the renormalization group from IR to UV scales in discrete steps $\mu \to 2\mu$ to match lattice calculations with perturbation theory. In this respect, even assuming that the $Q=0$ sector is correctly sampled by a topologically frozen algorithm,  the issue could be raised as to whether the $Q=0$ projection leads to a proper renormalization scheme, and thus to a correct step-scaling sequence.

In this paper, we address these issues with a novel numerical technique designed to solve topological freezing, the \emph{Parallel Tempering on Boundary Conditions} (PTBC) algorithm. This algorithm, originally proposed for $2d$ $\mathrm{CP}^{N-1}$ models by M.~Hasenbusch~\cite{Hasenbusch:2017unr}, and implemented for $4d$ $\SU(N)$ Yang--Mills theories too~\cite{Bonanno:2020hht}, has been widely employed in the last few years to improve the state of the art of the lattice studies of several topological and non-topological quantities, thanks to the impressive reduction of the auto-correlation time of the topological charge it allows to achieve~\cite{Berni:2019bch,Bonanno:2020hht,Bonanno:2022yjr,Bonanno:2022hmz,Bonanno:2023hhp,Bonanno:2024ggk} (see also~\cite{Bonanno:2024udh,Abbott:2024mix} for recent applications of the tempering on boundary conditions with non-equilibrium simulations and normalizing flows). Given that PTBC is a well-established method to avoid topological freezing, it is a natural choice to accurately check the effects of the $Q=0$-projection.

In a few words, the PTBC algorithm consists in the simulation of several replicas of the lattice, differing among themselves for the boundary conditions imposed on the gauge links in a small sub-lattice. These interpolate among Open Boundary Conditions (OBCs) and Periodic Boundary Conditions (PBCs). During the Monte Carlo, all lattice replicas get updated simultaneously and independently, and swaps of gauge configurations among different replicas are proposed. The idea is that a gauge configuration, thanks to the swaps, performs a random walk among the replicas, experiencing different boundary conditions. Since simulations with open boundaries have much smaller auto-correlation times~\cite{Luscher:2011kk,Luscher:2012av}, the decorrelation of topological charge gets ``transferred'' to the replica with PBCs, where all measures are performed. This last point is a crucial ingredient of the PTBC algorithm, as it allows to circumvent the unphysical effects introduced by OBCs; indeed, these require to stay sufficiently far from the boundaries, thus making it harder to keep finite-size effects under control.

The goal of the present investigation is twofold. First, we compare the values of the projected coupling obtained with a standard and the PTBC algorithms. The aim is to explicitly verify whether the former is able to correctly sample gauge configurations within the $Q=0$ sector regardless of topological freezing, and thus whether a bias due to ergodicity problems is present after projection. Some preliminary results about this point were presented at the 2023 Lattice conference~\cite{DasilvaGolan:2023cjw}. Secondly, we aim at verifying that the two definitions of the coupling, projected and non-projected, lead to the same flow of the renormalization scale in the step-scaling sequence. This would imply that topological projection defines a proper renormalization scheme.

As explained in detail in the following section, the outcomes of these two tests are sufficient to predict whether or not the projected and the non-projected coupling will eventually lead to the same results for the $\Lambda$-parameter, without the need to perform its whole computation. Our strategy, thus, has the advantage of disentangling the possible systematics introduced by topological freezing and topological projection from other sources of error. We anticipate that our results fully support the $Q=0$ projection.

This manuscript is organized as follows. In Sec.~\ref{sec:strategy} we explain in detail our strategy to check the impact of topological freezing and topological projection on the determination of $\Lambda$. In Sec.~\ref{sec:setup} we present our numerical setup, describing how we implemented the PTBC algorithm in combination with the twisted volume-reduced setup of Ref.~\cite{Bribian:2021cmg}, and the techniques employed to compute the strong coupling via the gradient flow. In Sec.~\ref{sec:res} we discuss our numerical results. Finally, in Sec.~\ref{sec:conclu} we draw our conclusions.

\section{The $\Lambda$-parameter and the step-scaling method}\label{sec:strategy}

Before explaining the strategy we followed to check the impact of topological projection on the determination of the running coupling and the $\Lambda$-parameter, we recall the definition of this quantity, and how it is computed on the lattice using the so-called \emph{step-scaling} method~\cite{Luscher:1991wu}.

The Gell-Mann--Low $\beta$-function defined in the regularization scheme $\s$,
\beq\label{eq:beta_func_def}
\beta_{\s}(\lambda_{\s}) &\equiv& \frac{d \lambda_{\s}(\mu)}{d \log(\mu^2)}  \nonumber \\
&\underset{\lambda_{\s} \,\to \, 0}{\sim}& - \lambda_{\s}^2\left(b_0 + b_1 \lambda_{\s} + b_2^{(\s)} \lambda_{\s}^2 + \dots\right),
\eeq
defines a first-order differential equation which expresses the running of the renormalized $\SU(N)$ 't Hooft coupling $\lambda_{\s}(\mu)\equiv N g^2_{\s}(\mu)$, and admits a perturbative expansion which is universal (i.e., scheme-independent) up to the 2-loop order. The equation in~\eqref{eq:beta_func_def} can be exactly integrated, and the scheme-dependent, renormalization-group invariant $\Lambda$-parameter is its related integration constant:
\beq\label{eq:lambda_def}
\frac{\Lambda_\s}{\mu} &=& [b_0\lambda_{\s}(\mu)]^{-\frac{b_1}{2b_0^2}} e^{-\frac{1}{2b_0\lambda_{\s}(\mu)}} \times \nonumber \\
\exp&&\left\{-\int_0^{\lambda_{\s}(\mu)} dx\left( \frac{1}{2\beta_{\s}(x)} + \frac{1}{2b_0x^2} - \frac{b_1}{2b_0^2x} \right)\right\}.
\eeq
Introducing two generic scales $\mu_1$ and $\mu_2$, the following exact relation holds:
\beq\label{eq:lambda_step_scal}
\frac{\Lambda_{\s}}{\mu_1} = \frac{\Lambda_{\s}}{\mu_2} \exp\left\{-\int_{\lambda_{\s}(\mu_2)}^{\lambda_{\s}(\mu_1)} \frac{dx}{2\beta_{\s}(x)} \right\}.
\eeq 
Take $\mu_1=\mu_\had$ and $\mu_2=\mu_\pt$ in the non-perturbative and in the perturbative regimes respectively. The scale $\mu_{\had}$ should be chosen such that it is possible to accurately determine from lattice simulations both $\lambda_{\s}(\mu_{\had})$ and $\mu_{\had}$ in physical units. Now suppose some transformation of the parameters of the simulations allows to determine $\lambda_{\s}(2\mu)$ from $\lambda_{\s}(\mu)$. Then, recursively iterating this \emph{step-scaling}~\cite{Luscher:1991wu} transformation $k$ times, it is possible to obtain $\lambda_{\s}(\mu_{\pt})$ with $\mu_{\pt} = 2^k\mu_{\had}$. This scale should be large enough that $\Lambda_\s/\mu_\pt$ can be evaluated with some perturbative truncation of Eq.~\eqref{eq:lambda_def}. Thanks to step-scaling, the exponential factor appearing in Eq.~\eqref{eq:lambda_step_scal} simply becomes
\beq\label{eq:step_scaling_factor}
&&\exp\left\{-\int_{\lambda_{\s}(\mu_\pt)}^{\lambda_{\s}(\mu_\had)} \frac{dx}{2\beta_{\s}(x)} \right\} = \nonumber \\ 
&& \exp\left\{-\int_{\mu_\pt}^{\mu_\had} d\log(\mu)\right\} = \frac{\mu_{\pt}}{\mu_{\had}} = 2^k.
\eeq
Then the $\Lambda$-parameter is eventually obtained as:
\beq\label{eq:lambda_with_step_scaling}
\Lambda_\s = \left(\frac{\Lambda_s}{\mu_\pt}\right)\Bigg\vert_{\pt} 2^k \mu_{\had}.
\eeq

As described above, the step-scaling procedure requires to compute the coupling $\lambda_\s(\mu_i= 2^i\mu_{\had})$ at each step $i$ up to the value $\lambda_\s(\mu_\pt)$ where the matching to perturbation theory is done. In light of the typical energy scales that can be reached on the lattice, it is crucial to perform a non-perturbative calculation of the running coupling over a substantial range of values, extending to sufficiently small couplings, in order to ascertain the actual size of the perturbative corrections and to ensure the reliability of the $\lambda_{\rm s}(\mu_{\rm pt}) \to 0$ limit.

Moreover, the determination of the step-scaling sequence is more conveniently done in terms of the so-called \emph{step-scaling function}:
\be
\sigma_\s(u) = \lambda_\s (\mu/2) \Big|_{\lambda_\s (\mu) \,= \, u}\,,
\ee
which measures the change of the coupling when the renormalization scale is divided by a factor of two. Then, $u_0 \equiv \lambda_\s (\mu_{\had})$ is chosen and the matching point with perturbation theory is calculated by repeated evaluations of the inverse of $\sigma_s(u)$:
\be
u_k  = \sigma_s^{-1}(u_{k-1}).
\ee

To run the scale $\mu$, it is convenient to work in a finite-volume renormalization scheme~\cite{Luscher:1991wu}, in which $\mu$ is linked to the physical volume of the system:
\be
\mu = \frac{1}{cl},
\ee
where $l$ is the physical extent and $c$ is an arbitrary $O(1)$ constant that defines the scheme. For a lattice simulation, $l = La$ where $L$ is the lattice extent and $a$ the lattice spacing. Then, the procedure to determine $\sigma_s(u)$ is the following:
\begin{itemize}
	\item \label{step1} Consider several simulations with different values of $L$. Define a Line of Constant Physics (LCP) by tuning the bare coupling for each lattice to achieve the same value of the renormalized coupling $u = \lambda_\s(\mu)$, where $\mu=1/(c l)$. Assuming a 1-to-1 correspondence between $\lambda_s$ and $\mu$, this is equivalent to tune the bare couplings to achieve a fixed value of $l$.
	
	\item \label{step2} For each lattice, keep the same value of the bare coupling but double the size, $L \,\rightarrow\, 2 L$, so that $\mu \,\rightarrow\, \mu/2$. Thus, the values of the renormalized coupling $\Sigma_\s(u,L)$ measured from these simulations are estimates of the step-scaling function, which can be obtained from a continuum extrapolation:
	\beq
	\sigma_\s(u)= \lim_{1/L \, \to \, 0} \Sigma_\s(u,L) 
	\eeq
\end{itemize}
Iterating these 2 steps from the perturbative to the low energy regimes, it is possible to determine the step-scaling sequence and use it to evaluate $\Lambda_\s/\mu_\had$.

We can now discuss the two possible issues associated with the topological projection of the renormalized coupling. The first one is algorithmic. Generally, it is assumed that topological freezing only affects the relative weights between different topological sectors. However, there is no proof that, in the presence of freezing, fluctuations in the topological charge density are correctly sampled within a given sector, i.e., that there is no loss of ergodicity at fixed topological charge. In the semiclassical domain, and for the zero sector, this would imply, for example, that instanton/anti-instanton pairs are correctly sampled, even if single instantons are not. Since topological freezing becomes more severe as the continuum limit is approached, a loss of ergodicity at fixed $Q$ could introduce a bias both in the tuning of the LCP determined by $u$ and in the continuum extrapolation of the step-scaling function. To analyze this issue, in Sec.~\ref{sec:res_LCP1} we compare the couplings defined in the $Q=0$ sector obtained using the standard and the PTBC algorithms.

The second issue concerns the topological projection itself, in particular whether it allows to define a legitimate renormalization scheme and how it could affect the determination of $\Lambda_s$. Before discussing this point, recall that the ratio between $\Lambda$-parameters in two different schemes can be obtained through a one-loop calculation in perturbation theory. Since in perturbation theory topology plays no role and only the $Q=0$ topological sector is relevant, the $Q=0$-projected and the non-projected coupling are exactly equal at one-loop order. Thus, the $\Lambda$-parameter should be the same in both schemes. Therefore, for the purpose of checking the impact of topological projection on $\Lambda_{\s}$, it is sufficient to examine only the first step of the procedure, from $\mu_\had$ to $2\mu_\had$, where topological fluctuations are less suppressed and the effects of the projection should be more important.~\footnote{We thank Alberto Ramos for pointing this out.}

The details of how we implemented such test are described in Sec.~\ref{sec:res}. A summary is the following:
\begin{enumerate}
	\item \label{pnt1}
	Tune the bare couplings on various lattices with different sizes $L$ to obtain the same target value $u^{(0)}_\tg$ for the projected renormalized coupling.
	\item \label{pnt2}
	From simulations with the same bare couplings and double sizes, $L\to 2L$, measure both the projected and non-projected couplings and extrapolate their continuum values: $\sigma_\s^{(0)}(u^{(0)}_\tg)$ and $\sigma_\s^{(\noproj)}(u^{(0)}_\tg)$\footnote{In this case, the argument $u^{(0)}_\tg$ of the step scaling function is used to indicate that the tuning of the LCP at $2\mu_\had$ is done in both cases in terms of the projected coupling.}.
	If everything is consistent, these two should define the same renormalization scale $\mu_\had$.
	\item \label{pnt3}
	Tune the bare coupling on the doubled lattices to obtain a constant value of the projected coupling equal to the previous continuum extrapolation $\sigma_\s^{(0)}(u^{(0)}_\tg)$. Then, determine the non-projected couplings. Given that lattice artefacts in the difference between projected and unprojected couplings are expected to be much smaller than those of each coupling separately, to a good extent we expect that on each lattice, and certainly in the continuum limit:
	\beq\label{eq:test}
	\lambda_\s^{(\noproj)}\Big|_{\sigma_\s^{(0)}(u^{(0)}_\tg)} =  \sigma_\s^{(\noproj)}(u^{(0)}_\tg)\, ,
	\eeq
	i.e., that also the non-projected coupling is constant and equal to the previously-determined value of the step-scaling function.
\end{enumerate}

\section{Numerical methods}\label{sec:setup}

In this section, we describe our numerical setup, namely, the lattice discretization adopted for the gauge action, the strong coupling and the topological charge, and the practical implementation of the PTBC algorithm we employed.

\subsection{Twisted volume reduction and twisted gradient flow coupling}

Concerning the lattice definition of the action and the observables, we follow the same numerical setup of Ref.~\cite{Bribian:2021cmg}, which we shortly review in this section.

We discretize the pure-gauge $\SU(3)$ theory using the Wilson plaquette action on a lattice with lattice spacing $a$, geometry $L^2 \times \tL^2$, with $\tL = L/N = L/3$, and Twisted Boundary Conditions (TBCs) along the short directions~\cite{tHooft:1979rtg,tHooft:1980kjq}. These two latter peculiar choices will be better justified shortly. In practice, the discretized action reads:
\beq\label{eq:lattice_action_TBC}
S_{\rm W}[U] = -N b\sum_{x,\mu>\nu} Z_{\mu\nu}^*(x)\Re\Tr \left[ P_{\mu\nu}(x)\right],
\eeq
where $b = 1/\lambda_L$ is the inverse bare 't Hooft coupling and $P_{\mu\nu}(x) = U_\mu(x) U_\nu(x+a\hat{\mu}) U_\mu^{\dag}(x+a\hat{\nu}) U^{\dag}_\nu(x)$ is the plaquette operator on site $x$ along the $(\mu,\nu)$ plane. Finally, the factor $Z_{\mu\nu}(x)$ is used to easily impose TBCs along the short plane, taken to be the $(\mu, \nu) = (1,2)$ plane in our case:
\beq
Z_{\mu\nu}(x) = Z_{\nu\mu}^*(x) =
\begin{cases}
	e^{i 2 \pi / 3},& \text{ if } (\mu,\nu)=(1,2) \text{ and }\\
	& \quad x_{\mu}=x_\nu=0,\\
	1, & \text{ elsewhere}.
\end{cases}
\eeq

The choice of a lattice with reduced extents along the twisted plane is rooted in the idea of \emph{twisted volume reduction} \cite{Gonzalez-Arroyo:1982hyq,Gonzalez-Arroyo:1982hwr,Gonzalez-Arroyo:2010omx} (see also Refs.~\cite{GarciaPerez:2014cmv,GarciaPerez:2020gnf} for reviews on the topic), a technique usually employed to study the large-$N$ limit of $\SU(N)$ gauge theories. Indeed, in the large-$N$ limit and under certain conditions satisfied by TBCs, Yang--Mills theories enjoy a dynamical equivalence between color and space-time degrees of freedom. This equivalence, known since the seminal paper of Eguchi and Kawai~\cite{PhysRevLett.48.1063}, leads to a volume-independence of the theory for $N=\infty$. This property, which strictly holds only in the large-$N$ limit, at finite $N$ and with TBCs allows an effective increase in the lattice size: $V_{\rm eff} = N^2 V$. Since our lattice has $V=L^2 \times \tL^2 = L^4 / N^2$, this means that $V_{\rm eff} = N^2 V = L^4$, i.e., we achieve the same dynamics of a standard hypercubic lattice with size $L$. Adopting TBCs also has other advantages: it allows an analytic expansion in the coupling in perturbation theory as opposed to PBCs~\cite{Luscher:1982ma}, and it is free of $O(a)$ effects presents, for instance, in the Schrödinger Functional scheme~\cite{Luscher:1992an}.

For what concerns the definition of the renormalized coupling, we make use of the gradient flow, a smoothing procedure that evolves the gauge fields according to the flow-time equations:
\beq
\partial_t B_\mu (x, t) = D_\nu F_{\nu \mu} (x, t), \quad B_\mu (x, t = 0) = A_\mu (x),
\eeq
where $D_\mu$ and $ F_{\mu \nu}$ stand for the covariant derivative and field strength tensor of the flowed fields, while $A_\mu(x)$ stands for the unflowed gauge field. The gradient flow introduces an additional length scale, the \emph{smoothing radius} $r_s = \sqrt{8 t}$, with $t$ the flow time in physical units. Thus, it is natural in this setup to identify the inverse of this scale as the energy scale $\mu$ of the running coupling. In turn, as already pointed out in Sec.~\ref{sec:strategy}, it is natural to choose this length scale as a fraction $c$ of the physical size of the lattice.  When combined with our asymmetric volume setup and twisted boundary conditions, the gradient flow leads to a particular scheme to define the coupling known as \emph{Twisted Gradient Flow} (TGF)~\cite{Ramos:2014kla,Bribian:2019ybc,Bribian:2021cmg}. In more concrete terms, the TGF renormalized coupling is defined in the continuum theory according to:
\beq\label{eq:coupling_continuum}
\lambda_{\TGF}\left(\mu = \frac{1}{c l}\right) = \mathcal{N}(c) \braket{t^2 E(t)}\Bigg\vert_{\sqrt{8 t} \, = \, c l}\,,
\eeq
where $E(t)$ is the energy density evaluated on the flowed fields,
\beq
E(t) = \frac{1}{2} \Tr \left \{F_{\mu \nu} (x, t)F_{\mu \nu} (x, t)\right\}\, ,
\eeq
and with $\mathcal{N}(c)$ a normalization factor given by
\beq
\mathcal{N}(c) &=& \frac{128 \pi^2}{3N \mathcal{A}(\pi c^2)}, \\
\mathcal{A}(x) &=& x^2\theta_3^2(0,ix)\left[\theta_3^2(0,ix) - \theta_3^2(0,ixN^2)\right],
\eeq
with $\theta_3(z,ix)=x^{-1/2}\sum_{m\in \mathbb{Z}}\exp(-\pi(m-z)^2/x)$ the Jacobi $\theta_3$ function. This ensures that, at lowest order of perturbation theory, $\lambda_{\TGF} = \lambda_{\MS} + O(\lambda_{\MS}^2)$. The value of $c$ can be freely chosen, and just amounts to define a particular regularization scheme; here we adopt $c=0.3$. Although we will not use it here, we also recall that the conversion factor between the $\Lambda$-parameters in the $\TGF$ and in the $\MS$ scheme is known~\cite{Bribian:2021cmg}.

As mentioned in the introduction, it is customary to address the issue of topological freezing by projecting the determination of the coupling into the sector of configurations with zero topological charge~\cite{Fritzsch:2013yxa}. Although only the projection to $Q=0$ leads to a coupling definition that matches the perturbative one at high energies, it is possible to generalize the projection to an arbitrary sector of charge $n$ as follows:
\beq
&&\lambda_{\TGF}^{(n)}\left(\mu = \frac{1}{c l}\right) = \frac{ 128 \pi^ 2 t^2 }{3 N  {\cal A}(\pi c^2)} \frac{\langle E\left(t\right)\delta(Q-n)\rangle}{\langle \delta(Q-n)\rangle} \Bigg\vert_{\sqrt{8 t} \, = \, c l} \\
&&Q = \frac{1}{32 \pi^2}\varepsilon_{\mu\nu\rho\sigma} \int d^4 x \, \Tr\left\{F_{\mu\nu}(x)F_{\rho\sigma}(x)\right\} \in \mathbb{Z} \, ,
\label{eq:lambdat}
\eeq
where $\delta(Q-n)$ stands for a $\delta$-function that restricts the calculation to configurations with topological charge $Q=n$. The non-projected coupling in Eq.~\eqref{eq:coupling_continuum}, averaged over all topological sectors, will be referred to in the following as $\lambda_{\TGF}^{(\noproj)}$.

On the lattice, we use the Wilson flow combined with twisted boundary conditions to determine the coupling, meaning that the gauge fields are evolved during the flow using exactly the action in Eq.~\eqref{eq:lattice_action_TBC}. As for the energy density, we used the clover-discretized energy density given by:
\beq
\!\!E_{\clov}(t) = \frac{1}{2} \Tr\left[C_{\mu\nu}(x,t)C_{\mu\nu}(x,t)\right],
\eeq
with $C_{\mu\nu}(x,t)$ the clover operator in the site $x$ along the $(\mu,\nu)$ plane,
\beq
C_{\mu\nu}(x,t) &=& \frac{1}{4}\Im [Z^*_{\mu\nu}(x) P_{\mu\nu}(x,t) \nonumber \\   
& +& Z^*_{\mu\nu}(x-a\hat{\nu}) P_{-\nu\mu}(x,t)  \nonumber \\   
& +& Z^*_{\mu\nu}(x-a\hat{\mu}) P_{\nu-\mu}(x,t) \nonumber \\   
& +& Z^*_{\mu\nu}(x-a\hat{\mu}-a\hat{\nu}) P_{-\mu-\nu}(x,t)],
\eeq
where $U_{-\mu}(x) = U_{\mu}^{\dagger}(x-a\hat{\mu})$. In order to eliminate the leading lattice artefacts in perturbation theory for the Wilson flow, we also take a discretized version of the normalization constant ${\cal N}$:
\beq
&&{\cal N}_L^{-1}(c, L) = \\
&&\frac{c^4}{128}\sum_{\mu\neq\nu} \sum_{q}^{'} e^{-\frac{1}{4}c^2 L^2\hat{q}^2}\frac{1}{\hat{q}^2}\sin^2(q_\nu)\cos^2(q_\mu/2),
\nonumber
\label{latt_norm}
\eeq
where $\hat q_\mu =2\sin(q_\mu/2)$ stands for
the lattice momentum, with $q_\mu = 2 \pi n_\mu /L$, $n_\mu = 0, \cdots, L-1$,
and with the prime in the sum denoting the exclusion of momenta with both components in the twisted plane satisfying $L q_i \propto 2 N \pi = 6\pi$. 

The TGF technique will be also used to define the topological charge on the lattice. In particular, we will adopt the simplest parity-defined clover discretization,
\beq
Q_{\clov}(t) = \frac{1}{32\pi^2}\sum_{x,\mu\nu\rho\sigma}\varepsilon_{\mu\nu\rho\sigma}\Tr\left[C_{\mu\nu}(x,t)C_{\rho\sigma}(x,t)\right],
\eeq
and define our physical topological charge and topological susceptibility after the flow, at the same flow time employed to define the coupling:
\beq
Q = Q_{\clov}(\sqrt{8t}=c l), \qquad \qquad a^4\chi = \frac{\braket{Q^2}}{\tL^2 L^2}.
\eeq
In our simulations, this amount of flow turned out to be in all cases well within the observed plateau in $Q$ as a function of $t$ for large enough flow times, and the flowed clover charge at $\sqrt{8t}=c l$ always turned out to be extremely close to an integer number. Therefore, since the flowed definition of the topological charge has a well-defined continuum limit, one can define the projected coupling onto the topological sector $Q=n$ as follows:
\beq
\lambda_{\TGF}^{(n)}\left(\mu = \frac{1}{c l}\right) = \mathcal{N}_L(c, L) \frac{\braket{t^2 E_{\clov}(t) \hat \delta(Q-n)}}{\braket{\hat \delta(Q-n)}}\Bigg\vert_{\sqrt{8 t} \, = \, c l}
\eeq
where
\beq
\hat \delta(Q-n) =
\begin{cases}
	1, &\vert Q - n \vert < 0.5\\
	0, &\text{otherwise}.
\end{cases}
\eeq

\subsection{The PTBC algorithm in the presence of twisted boundary conditions}

In order to circumvent topological freezing for fine lattice spacings, we adopt the $\SU(N)$ PTBC algorithm of Ref.~\cite{Bonanno:2020hht}, which can be easily generalized to the current setup with TBCs. In practice, we consider $N_r$ replicas of the lattice, each one differing for the boundary conditions imposed on a small sub-region, which in the following will be addressed as the \emph{defect}. 
We choose the defect $D$ to be an $L_d \times L_d \times L_d$ spatial cube, placed on the time boundary $x_0 = L-1$. Moreover, only links that cross $D$ orthogonally are affected by its presence. This way, the tempering will always affect links that, in the physical replica (i.e., the one on which observables are computed), enjoy PBCs. Concerning the unphysical replicas, the idea is to choose their boundary conditions on the defect in such a way to interpolate between PBCs and OBCs. This is achieved by taking the action of the replica $r$ of the form:
\begin{equation}
	S_{\rm W}^{\left(c(r)\right)}[U_r] = -Nb\sum_{x,\mu>\nu} K_{\mu\nu}^{\left(c(r)\right)}(x) Z_{\mu\nu}^*(x)\Re\Tr \left[ P^{(r)}_{\mu\nu}(x)\right],
\end{equation}
where $U_r$ denotes the gauge links of the replica $r$. The factor $K_{\mu\nu}^{\left(c(r)\right)}(x)$, used to change the boundary conditions on the defect similarly to the twist factor $Z_{\mu\nu}(x)$, is:
\beq
\begin{aligned}
	K_{\mu\nu}^{\left(c(r)\right)}(x) &\equiv \,\, K_{\mu}^{\left(c(r)\right)}(x) K_{\nu}^{\left(c(r)\right)}(x+a\hat{\mu}) \\
	&\qquad\times \,\, K_\mu^{\left(c(r)\right)}(x+a\hat{\nu})K_\nu^{\left(c(r)\right)}(x),
\end{aligned}
\eeq
\beq
K_{\mu}^{\left(c(r)\right)}(x) &\equiv&
\begin{cases}
	c(r), \qquad &\mu=0, \,\quad x_0=L-1, \, \text{ and}\\
	&\quad 0 \le x_1,x_2,x_3 < L_d\\
	1,    \qquad &\text{elsewhere},
\end{cases}
\eeq
with $0\le c(r) \le 1$, where the edge cases $0$ and $1$ correspond, respectively, to open and periodic boundaries. In the following, all observables will be computed in the physical replica $r=0$ with $c(r=0)=1$.

For what concerns the Monte Carlo PTBC sampling algorithm, each replica is updated simultaneously and independently by performing 1 lattice sweep of the standard local heat-bath algorithm~\cite{Creutz:1980zw,Kennedy:1985nu}, followed by $n_{\ov}$ lattice sweeps of the standard local over-relaxation algorithm~\cite{Creutz:1987xi}. Then swaps among two adjacent replicas $(r,s=r+1)$ are proposed, and accepted via a standard Metropolis step:
\beq
p(r,s) = \min\left\{1, e^{-\Delta S^{(r,s)}_{\rm swap}}\right\},
\eeq
\beq
\begin{aligned}
	\Delta S^{(r,s)}_{\rm swap} =& \quad S_{\rm W}^{\left(c(r)\right)}[U_s]+S_{\rm W}^{\left(c(s)\right)}[U_r] \\
	& -S_{\rm W}^{\left(c(r)\right)}[U_r]-S_{\rm W}^{\left(c(s)\right)}[U_s].
\end{aligned}
\eeq
Note that, for the purpose of calculating $\Delta S^{(r,s)}_{\rm swap}$, one does not need to iterate over the whole lattice, as the only non-vanishing contributions to it come from the links found at most at a one lattice spacing distance from the defect. Given that the optimal setup is achieved when the mean acceptances $\mathcal{P}_r \equiv \braket{p(r,r+1)}$ are roughly constant, so that a given configuration can perform a sort of random walk among different replicas, we performed short test runs to tune the $c(r)$ tempering parameters in order to achieve $\mathcal{P}_r \approx \mathcal{P} \approx 20 \%$. With this choice, the number of replicas necessary to achieve a given constant mean acceptance $\mathcal{P}$ becomes just a function of the defect size in lattice units $L_d$.

Between two full updating sweeps involving the whole lattice, we performed several hierarchical updates on small sub-lattices centered around the defect, in order to update more frequently the links with tempered boundary conditions. This is done to improve the efficiency of the algorithm, as this is the region where new topological excitations are more likely to be created/destroyed. Moreover, after each swap is proposed, we translate the links of the periodic replica by one lattice spacing in a random direction, moving also consistently the position of the twisted plaquettes. This step is done to effectively move the position of the defect around the lattice, which is expected to improve the efficiency of the algorithm, as in this way topological excitations are created/destroyed in different space-time points. For further technical details regarding hierarchical updates, we refer the reader to the original papers~\cite{Hasenbusch:2017unr,Bonanno:2020hht}.

In a few words, given that the numerical effort required by hierarchical udpates, translations and swaps is negligible compared to the full sweeps of the lattice, one full parallel tempering updating step requires a numerical effort which is of the order of $N_r \times n_{\ov}$. This observation will be crucial to compare the efficiency of this algorithm with the standard one.

\section{Results}\label{sec:res}

In this section, we discuss the impact of topology on the determination of the strong coupling by comparing results obtained using the standard and the PTBC algorithms. Following the strategy described in Sec.~\ref{sec:strategy}, we perform the first step in the scaling sequence connecting $\mu_\had$ with $2\mu_\had$ using both the projected and non-projected couplings, as determined with PTBC and the standard algorithms. This first step reproduces the one used in Ref.~\cite{Bribian:2021cmg} to determine the $\Lambda$-parameter. The idea is the following:

\begin{table*}[!t]
	\begin{center}
		\begin{tabular}{|c|c|c|c|c|c|c|c|}
			\hline
			\multicolumn{8}{|c|}{}\\[-1em]
			\multicolumn{8}{|c|}{LCP1 ($l=0.55$~fm)}\\
			\hline
			&&&&&&&\\[-1em]
			$L$ & $18 \times b$ & $N_r$ & $L_d$ & $n^{(\rm PTBC)}_{\ov}$ & $n^{(\rm std)}_{\ov}$ & \makecell{Statistics\\(PTBC)} & \makecell{Statistics\\(Standard)} \\
			\hline
			12 & 6.4881 & 10 & 2 & 12 & 12 & 679850 & 90000 \\
			18 & 6.7790 & 17 & 3 & 12 & 18 & 150013 & 80000 \\ 
			24 & 7.0000 & 24 & 4 & 12 & 24 & 13388  & 23439 \\
			\hline
		\end{tabular}
		\begin{tabular}{|c|c|}
			\hline
			\multicolumn{2}{|c|}{}\\[-1em]
			\multicolumn{2}{|c|}{Scale setting}\\
			\hline
			&\\[-1em]
			$18 \times b$ & $a/\sqrt{t_0}$ \\
			\hline
			6.4881 & 0.2770(35)\\
			6.7790 & 0.1846(24)\\ 
			7.0000 & 0.1385(18)\\
			\hline
		\end{tabular}
	\end{center}
	\begin{center}
		\begin{tabular}{|c|c|c|c|c|c|c|c|}
			\hline
			\multicolumn{8}{|c|}{}\\[-1em]
			\multicolumn{8}{|c|}{LCP1 with doubled $L$}\\
			\hline
			&&&&&&&\\[-1em]
			$L$ & $18 \times b$ & $N_r$ & $L_d$ & $n^{(\rm PTBC)}_{\ov}$ & $n^{(\rm std)}_{\ov}$ & \makecell{Statistics\\(PTBC)} & \makecell{Statistics\\(Standard)} \\
			\hline
			24 & 6.4881 & 18 & 4 & 12 & 24 & 7783 & 10000 \\
			36 & 6.7790 & 34 & 6 & 12 & 36 & 5731 & 3203 \\ 
			48 & 7.0000 & 54 & 8 & 12 & 48 & 1805 & 2305 \\
			\hline
		\end{tabular}
	\end{center}
	\begin{center}
		\begin{tabular}{|c|c|c|c|c|c|c|c|}
			\hline
			\multicolumn{8}{|c|}{}\\[-1em]
			\multicolumn{8}{|c|}{LCP2 ($l=1.1$~fm)}\\
			\hline
			&&&&&&&\\[-1em]
			$L$ & $18 \times b$ & $N_r$ & $L_d$ & $n^{(\rm PTBC)}_{\ov}$ & $n^{(\rm std)}_{\ov}$ & \makecell{Statistics\\(PTBC)} & \makecell{Statistics\\(Standard)} \\
			\hline
			24 & 6.459 & 18 & 4 & 12 & 24 & 28787 & 61575 \\
			36 & 6.765 & 34 & 6 & 12 & 36 & 22234 & 46266 \\ 
			48 & 6.992 & 54 & 8 & 12 & 48 & 11658 & 25157 \\
			\hline
		\end{tabular}
	\end{center}
	\caption{Summary of simulation parameters, where the number of replicas $N_r$ and the defect size $L_d$ only refer to runs with the PTBC algorithm. The numbers $n^{(\rm PTBC)}_{\ov}$ and $n^{(\rm std)}_{\ov}$ refer to the number of over-relaxation lattice sweep per over-heat lattice sweep for, respectively, the PTBC and the standard algorithm. The scale setting was taken from Refs.~\cite{Luscher:2010iy,Knechtli:2017xgy,Giusti:2018cmp} or from a spline interpolation of data thereof. The defect size in lattice units $L_d$ was scaled in order to keep its length constant in physical units: $L_d/L = 1/6$. The number of replicas was scaled as a function of $L_d$ in order to achieve in all cases an almost uniform swap acceptance rate of $\sim 20 \%$ among adjacent replicas. Statistics for both algorithms refers to the total number of measures, collected every $n_{\rm ov}/2$  updating steps.}
	\label{tab:summary_simulations}
\end{table*}

\begin{enumerate}[label=(\Alph*)]
	\vspace*{0.1\baselineskip}
	\item\label{stepA} First, in simulations performed with the standard algorithm, we tune the bare couplings $b$ to have an approximately fixed value of the $Q=0$ projected coupling $u^{(0)}_\tg \equiv \lambda_{\TGF}^{(0)}(2\mu_\had)$ on $L=12,18,24$ lattices.\footnote{It is possible to account for a small mismatch in the target couplings later on, as described in Sec.~\ref{sec:res_LCP2}.} If everything is consistent, this set would correspond to an LCP with approximately fixed physical volume $l=a L$, which we dub LCP1. Then, we perform simulations for the same values of $b$ but on doubled lattices with sizes $L = 24, 36, 48$. The corresponding projected couplings, extrapolated to the continuum limit, give the step-scaling function $\sigma_{\TGF}^{(0)}(u^{(0)}_\tg)$. The results of this step are discussed in Sec.~\ref{sec:res_LCP1}.
	\vspace*{\baselineskip}
	\item\label{stepB} Finally, on lattices with $L=24,36,48$, we determine the bare couplings $b$ for which the $Q=0$ projected coupling takes the value $\lambda_{\TGF}^{(0)} = \sigma_{\TGF}^{(0)}(u^{(0)}_\tg)$ determined in~\ref{stepA}; details on how to do the tuning are provided in Ref.~\cite{Bribian:2021cmg}. This last set of simulation points should correspond to an LCP with fixed physical volume $l = aL = 1/(c \mu_\had)$, which we dub as LCP2. These simulations, whose results will be used to assess the impact of topological projection on the step-scaling sequence $\mu_\had \to 2 \mu_\had$, are discussed in Sec.~\ref{sec:res_LCP2}.
	
\end{enumerate}

The simulations outlined in~\ref{stepA} and~\ref{stepB} will be also performed with the PTBC algorithm using the same lattice sizes and bare couplings. All simulation parameters are summarized in Tab.~\ref{tab:summary_simulations}, where we also report the employed scale setting, which in Ref.~\cite{Bribian:2021cmg} was done using the standard gradient flow scale $t_0$~\cite{Luscher:2010iy}. For brevity, we moved the list of obtained lattice couplings to App.~\ref{app:raw_data}.

\begin{figure}[!t]
	\centering
	\resizebox{0.43\textwidth}{!}{\includegraphics{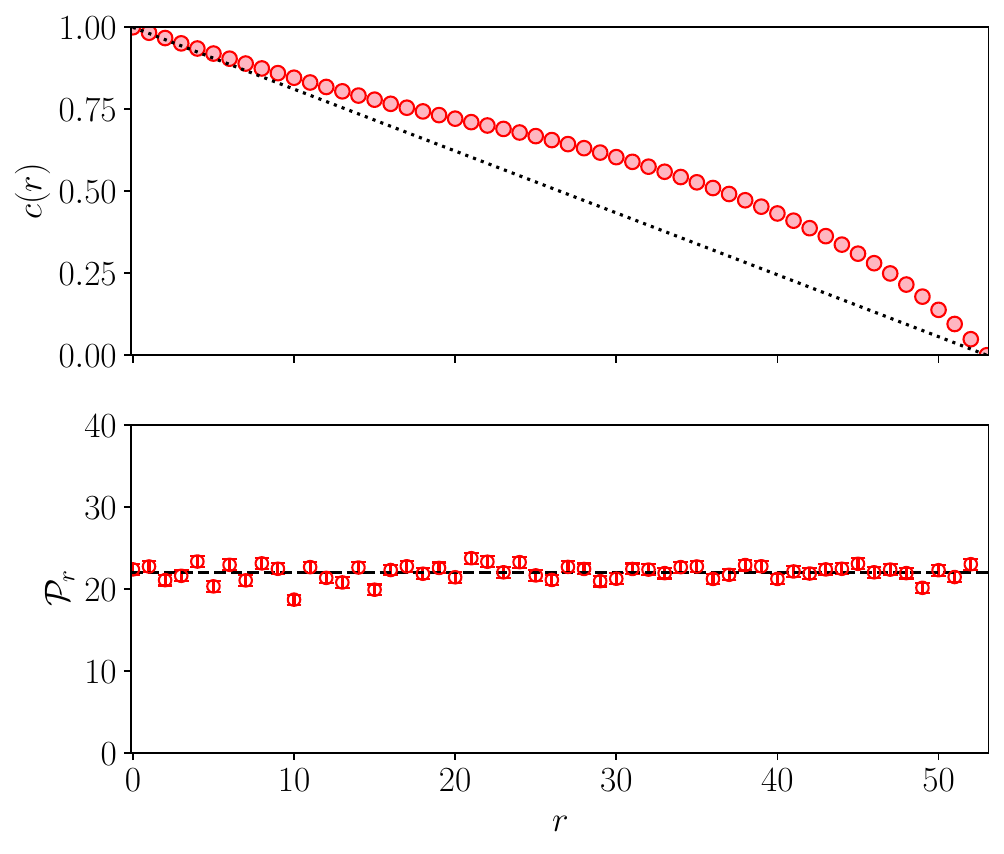}}
	\caption{This figure refers to the PTBC run with $L=48$, $18b = 6.992$, $N_r=48$, $L_d=8$. Top panel: tuned values of $c(r)$ compared with a simple linear behavior $c(r) = 1-r/(N_r-1)$. Bottom panel: corresponding mean swap acceptance rates $\mathcal{P}_r = \braket{p(r,r+1)}\approx 22(2)\%$.}
	\label{fig:swap_ex}
\end{figure}

Concerning simulations with parallel tempering, following the prescription advocated in the original references~\cite{Hasenbusch:2017unr,Bonanno:2020hht}, we kept the defect size fixed in physical units as we approached the continuum limit. This of course requires to scale the defect size $L_d$ as $1/a$. Since we also kept the mean acceptance swap rate fixed to $\approx 20\%$ for each adjacent replica couple (cf.~Fig.~\ref{fig:swap_ex}), the number of replicas $N_r$ is just a function of $L_d$, and is empirically found to scale approximately as $N_r \sim L_d^{3/2} \sim 1/a^{3/2}$, cf.~Tab.~\ref{tab:summary_simulations}, in agreement with the findings of Refs.~\cite{Hasenbusch:2017unr,Bonanno:2020hht}. As already pointed out in the previous section, the numerical cost of one parallel tempering updating step is of the order of $N_r \times n_{\ov}^{(\PTBC)}$, while the computational cost of one standard updating step is of the order of $n_{\ov}^{(\rm std)}$, where $n_{\ov}$ stands for the number of over-relaxation lattice sweeps per heat-bath lattice sweep. Thus, in the following, it will be more convenient to just compare the two algorithms expressing their Monte Carlo times in terms of a common scale, the number of lattice sweeps $n_{\rm sweeps} = N_r \times n_{\ov} \times n_{\rm steps}$, with $n_{\rm steps}$ the number of updating steps (of course $N_r=1$ for the standard algorithm).

\subsection{Impact of topological freezing on the $Q=0$ topology-projected strong coupling}\label{sec:res_LCP1}

We start our investigation by comparing the $Q=0$ projected couplings obtained with the standard and the PTBC algorithms in simulations with identical parameters.  The aim is to check whether or not the standard algorithm exhibits an ergodicity problem sampling the $Q=0$ sector. The comparison will be done for the LCP1 and the corresponding double lattices, as indicated in Tab.~\ref{tab:summary_simulations}.

Let us start from the LCP1, where the values of the bare couplings were chosen in Ref.~\cite{Bribian:2021cmg} to achieve an approximately constant value of the $Q=0$ projected coupling $\lambda_{\TGF}^{(0)} \approx 13.93(5)$.  This corresponds to an almost constant lattice size $l = aL \sim 0.55$ fm, and to an energy scale $\mu = 2\mu_\had = 1/(c l) = 1/(0.3 l) \sim 1.2$ GeV. We stress that, in this paper, we have increased the statistics used in Ref.~\cite{Bribian:2021cmg} for the lattice simulations of LCP1 performed with the standard algorithm; in addition, we have also repeated them using PTBC.

Since in the thermodynamic limit the topological susceptibility of the pure $\SU(3)$ gauge theory is $t_0^2 \chi = 6.67(4)\cdot 10^{-4}$~\cite{Ce:2015qha}, and since for our simulations $l/\sqrt{t_0} \simeq 3.32$, we can set the following very loose upper bound for these runs: $\braket{Q^2} \lesssim \tl^2 l^2 \chi = (l/\sqrt{t_0})^4 t_0^2\chi / N^2 \sim 0.009 \ll 1$. This upper bound, set with the infinite-volume result for $\chi$ and thus overestimated, shows that we can expect a tiny value of $\braket{Q^2}$. This means that topological fluctuations not only can be inhibited by topological freezing but are also strongly suppressed by the smallness of the volume. More precisely, assuming $P_0 \gg P_1 \gg P_2 \gg ...$, where $P_n$ is the probability of visiting the topological sector with $Q=n$, and using that $P_{-n} = P_n$, the following approximation holds:
\beq
\braket{Q^2} \simeq \frac{2P_1 + \dots}{P_0 + 2P_1 + \dots} \simeq \frac{2 P_1}{P_0} \lesssim 9 \cdot 10^{-3} .
\eeq
Thus, we can expect the probability of visiting the topological sector with $\vert Q \vert = 1$ to be at least two orders of magnitude smaller than the probability of visiting $Q=0$. This problem is well known in the finite-temperature QCD literature, as sufficiently above the QCD chiral crossover $T_c\simeq 155$ MeV the topological susceptibility is rapidly suppressed as $\chi\sim (T/T_c)^{-8}$~\cite{RevModPhys.53.43,Borsanyi:2015cka,Petreczky:2016vrs,Borsanyi:2016ksw,Jahn:2018dke,Bonati:2018blm,Lombardo:2020bvn,Borsanyi:2021gqg,Athenodorou:2022aay}: on typical volumes and for sufficiently large temperatures, $\braket{Q^2} = V \chi \ll 1$, as in the present case. 

Being the damping of topological fluctuations in this case mainly due to a physical effect, not to topological freezing, we expect a small number of topological fluctuations even when running with the parallel tempering. Moreover, one also expects $\lambda_{\TGF}^{(0)} \simeq \lambda_{\TGF}^{(\noproj)}$, given that the contribution from higher-charge sectors is highly suppressed.\footnote{Actually, to definitively conclude that $\lambda_{\TGF}^{(0)} \simeq \lambda_{\TGF}^{(\noproj)}$, in principle one should also check that $\lambda_{\TGF}^{(0)}/\lambda_{\TGF}^{(1)} \gg P_1/P_0$, which we are currently unable to do with the current setup for these simulation points with small volumes, as we cannot reliably measure $\lambda_{\TGF}^{(1)}$. To this end, one should employ one of the several strategies that have been devised in the literature to sample rare events, such as the multicanonic algorithm~\cite{Jahn:2018dke,Bonati:2018blm,Athenodorou:2022aay,Bonanno:2022dru}, or the density of states method~\cite{Borsanyi:2021gqg,Lucini:2023irm}.} This is perfectly reasonable, as in our setup smaller volumes mean larger energy scales, and closer to the perturbative regime we expect the $Q=0$ sector to largely dominate over the others.

\begin{figure*}[!t]
	\centering
	\resizebox{0.75\textwidth}{!}{\includegraphics{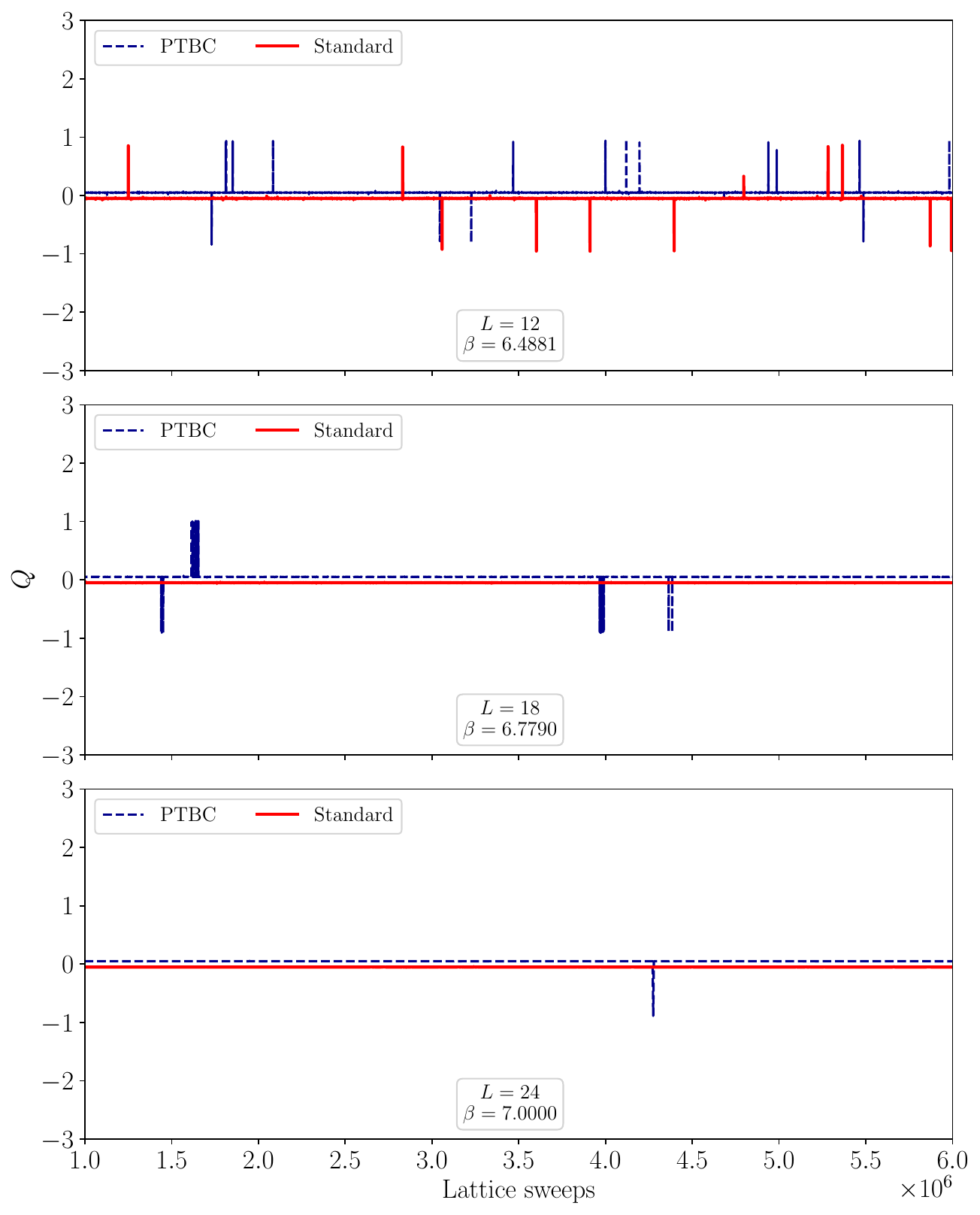}}
	\caption{Comparison of the Monte Carlo evolutions of the topological charge obtained with the standard and the PTBC algorithm for the 3 simulations points corresponding to a LCP with fixed lattice size $l \simeq 0.55$~fm and fixed projected coupling $\lambda_{\TGF}^{(0)} \approx 13.93(5)$. For readability, only a fraction of the total statistics is shown. In both cases, the horizontal Monte Carlo time was expressed in units of lattice sweeps in order to make a fair comparison among the two algorithms. This means that the number of updating steps in both cases was multiplied by the number of over-relaxation sweeps per heat-bath sweep, $n_{\ov}$, and, in the case of PTBC, also by the number of replicas $N_r$.}
	\label{fig:story_Q_LCP1}
\end{figure*}

\begin{figure}[!t]
	\centering
	\resizebox{0.45\textwidth}{!}{\includegraphics{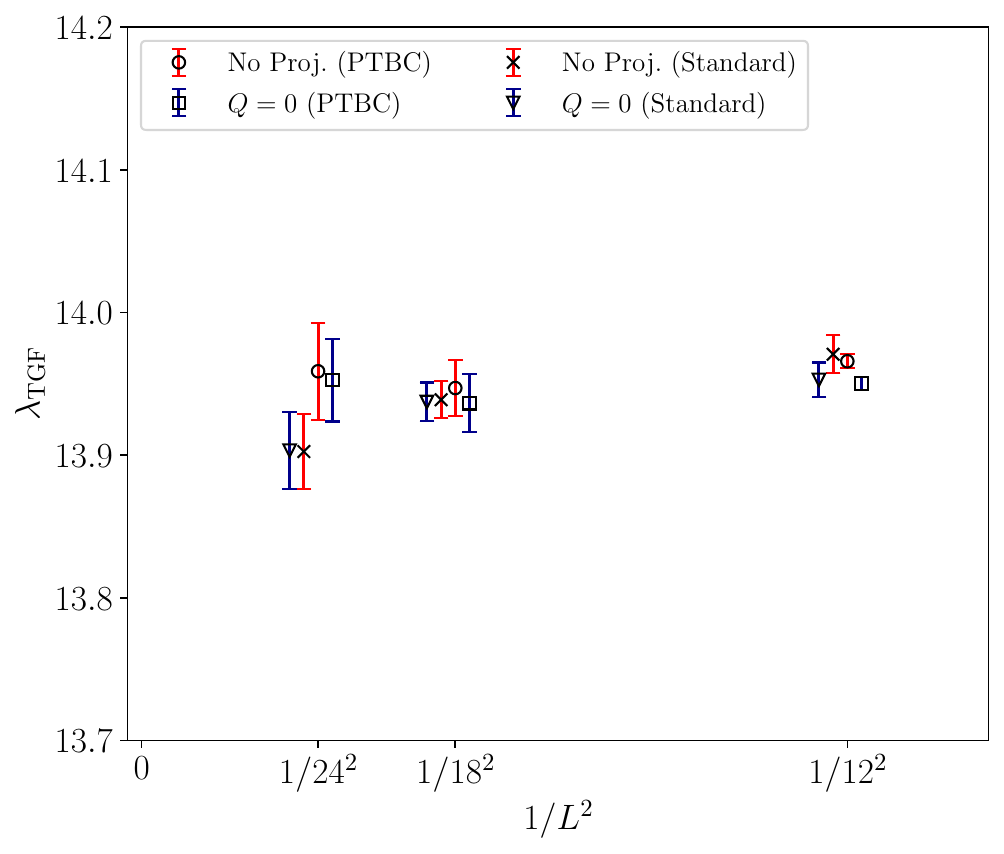}}
	\caption{Comparison of the $Q=0$ projected and non-projected couplings obtained with the standard algorithm in Ref.~\cite{Bribian:2021cmg} with those obtained in the present work with the PTBC algorithm for the 3 simulations points corresponding to a LCP with $l=0.55$~fm, tuned to achieve a constant value of $\lambda_{\TGF}^{(0)}$.}
	\label{fig:lambda_LCP1}
\end{figure}

These expectations are confirmed by our results for the Monte Carlo evolution of the flowed lattice topological charge, a part of which is shown in Fig.~\ref{fig:story_Q_LCP1}, and of the coupling, shown in Fig.~\ref{fig:lambda_LCP1}. As can be observed, the number of $Q\ne 0$ events remains rather small even in the Monte Carlo histories of the topological charge obtained with the PTBC algorithm, as the suppression of topological fluctuations has a physical origin, and the $Q=0$ sector dominates the actual topological charge distribution. This of course means that no difference can be appreciated between the $Q=0$ projected and the non-projected couplings. As a matter of fact, in all cases we observe at most differences at the level of one standard deviation within the per mil accuracy with which we have determined the coupling. We thus conclude that, for the purpose of calibrating the LCP1, both the projected and the non-projected coupling lead to perfectly consistent results.

\begin{figure}[!t]
	\centering
	\resizebox{0.45\textwidth}{!}{\includegraphics{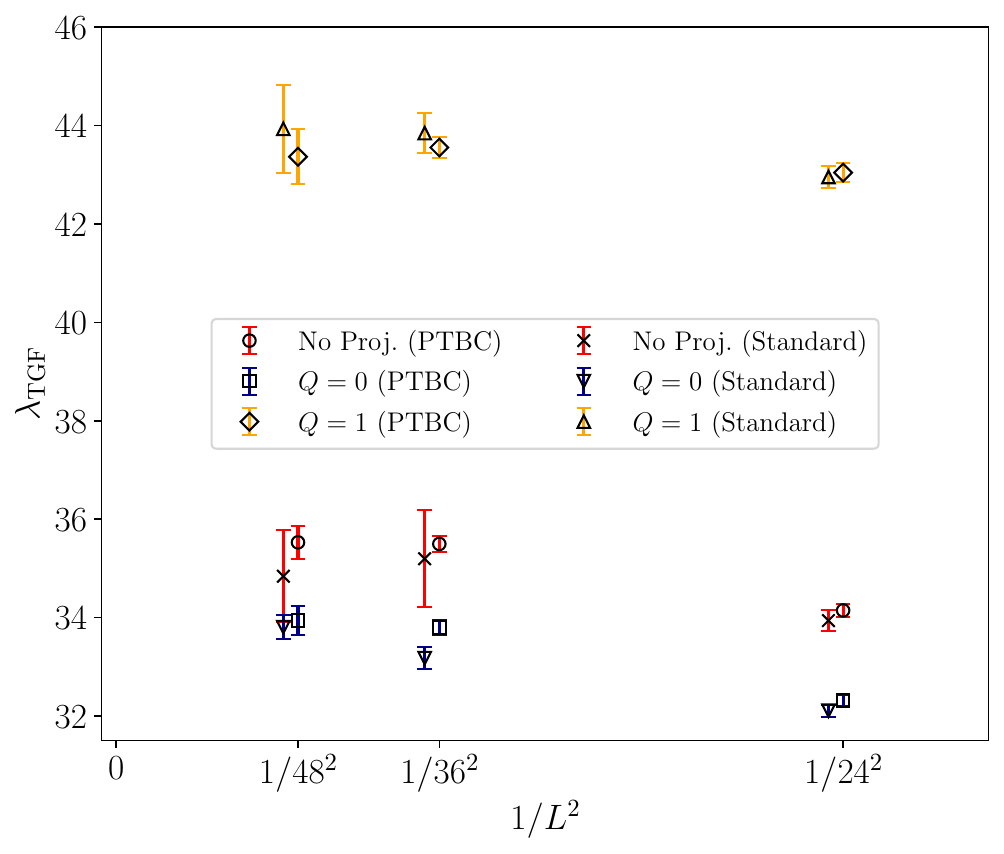}}
	\caption{Comparison of the $Q=0$ projected and non-projected couplings obtained with the standard algorithm and the PTBC algorithm for the 3 simulations points corresponding to the same bare couplings of the LCP with $l=0.55$~fm, but on lattices with doubled sizes.}
	\label{fig:lambda_vol2}
\end{figure}

\begin{figure*}[!t]
	\centering
	\resizebox{0.75\textwidth}{!}{\includegraphics{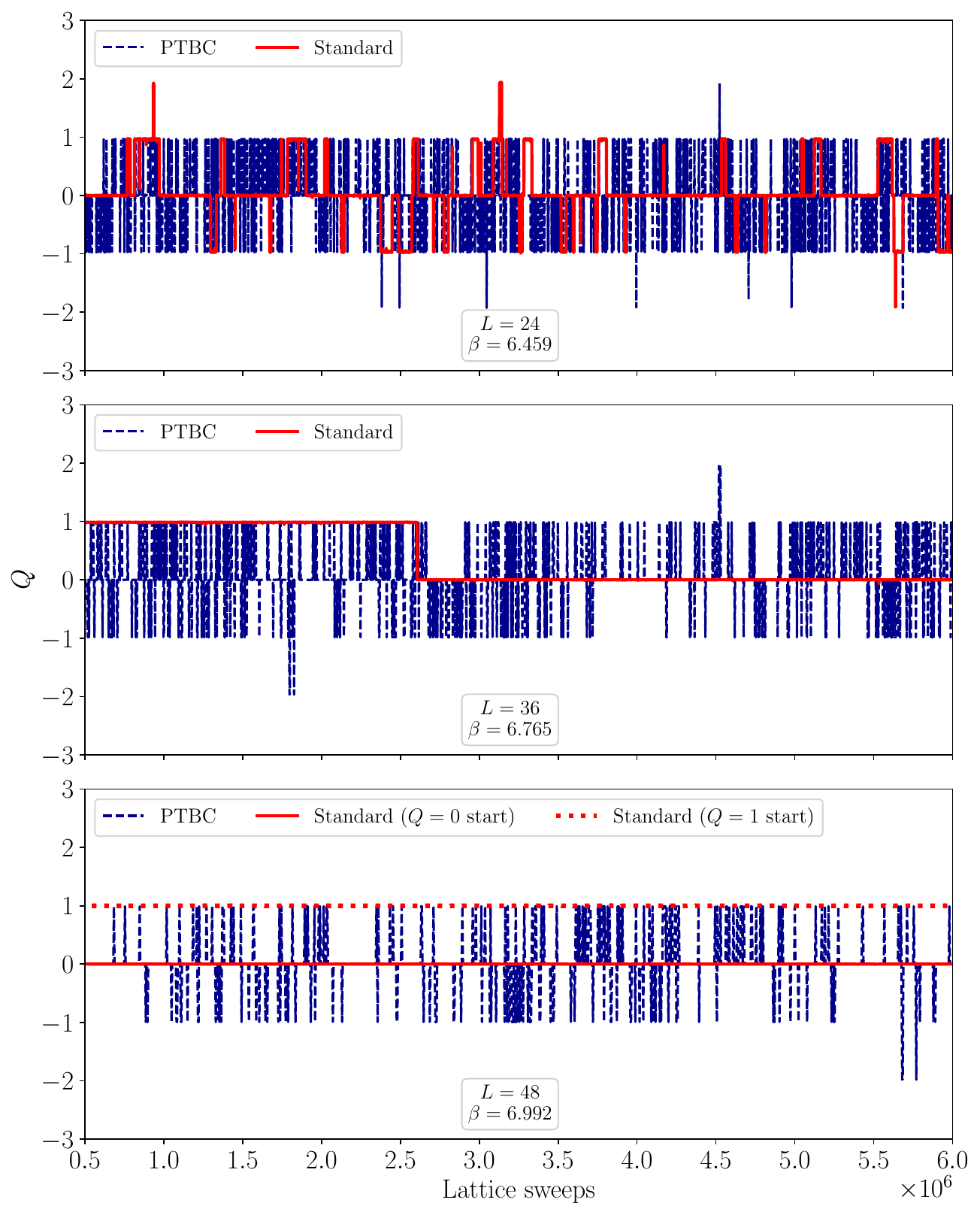}}
	\caption{Comparison of the Monte Carlo evolutions of the topological charge obtained with the standard and the PTBC algorithm for the 3 simulations points corresponding to a LCP with $l=1.1$~fm. For readability, only a fraction of the total statistics is shown. In both cases, the horizontal Monte Carlo time was expressed in units of lattice sweeps in order to make a fair comparison among the two algorithms. This means that the number of updating steps in both cases was multiplied by the number of over-relaxation sweeps per heat-bath sweep, $n_{\ov}$, and, in the case of PTBC, also by the number of replicas $N_r$.}
	\label{fig:comp_Q_LCP2}
\end{figure*}

We now move to the computation of the lattice step-scaling function $\Sigma_{\TGF} (u, L)$ using the same bare couplings of LCP1 on the doubled lattices. Given that we have now doubled the lattice sizes, we expect $\braket{Q^2} \sim O(0.1)$, thus we foresee topological fluctuations to start to become important. This in turn implies that $\lambda_{\TGF}^{(\noproj)}$ and $\lambda_{\TGF}^{(0)}$ now will differ sizably. Results for the projected and the non-projected couplings, obtained both with the PTBC and the standard algorithms, are shown in Fig.~\ref{fig:lambda_vol2}. As expected, we now observe a sizeable difference between the projected and the non-projected couplings, due to the contribution of higher-charge sectors, which are now much less suppressed. However, concerning the projected couplings, we observe that the results obtained with the standard algorithms for the $Q=0$ and $Q=1$ sectors are in perfect agreement with those obtained with PTBC, as at most we observe $1-2$ standard deviation differences within our per mil accuracy. This is a very non-trivial check that projection works even in the presence of severe topological freezing.  

Finally, let us conclude our discussion by comparing the performances of the standard and the PTBC algorithms. For that we will use long-run simulations performed on the lattice corresponding to the LCP2 in Tab.~\ref{tab:summary_simulations}. While the standard algorithm exhibits significant topological freezing, especially at the two finest lattice spacings explored, the PTBC one allows to achieve an impressive improvement in the observed number of topological fluctuations at fixed parameters. Such improvement can be clearly seen by inspecting Fig.~\ref{fig:comp_Q_LCP2}, where we compare the Monte Carlo evolutions of $Q$ obtained with the two algorithms, after expressing the Monte Carlo time in the same units in both cases. Note that for the finest lattice spacing we observed no fluctuations of $Q$ with the standard algorithm, and two independent Monte Carlo histories started from configurations with $Q=0$ and $Q=1$ both remained stuck in the initial topological sector.

\begin{figure*}[!t]
	\centering
	\resizebox{0.5\textwidth}{!}{\includegraphics{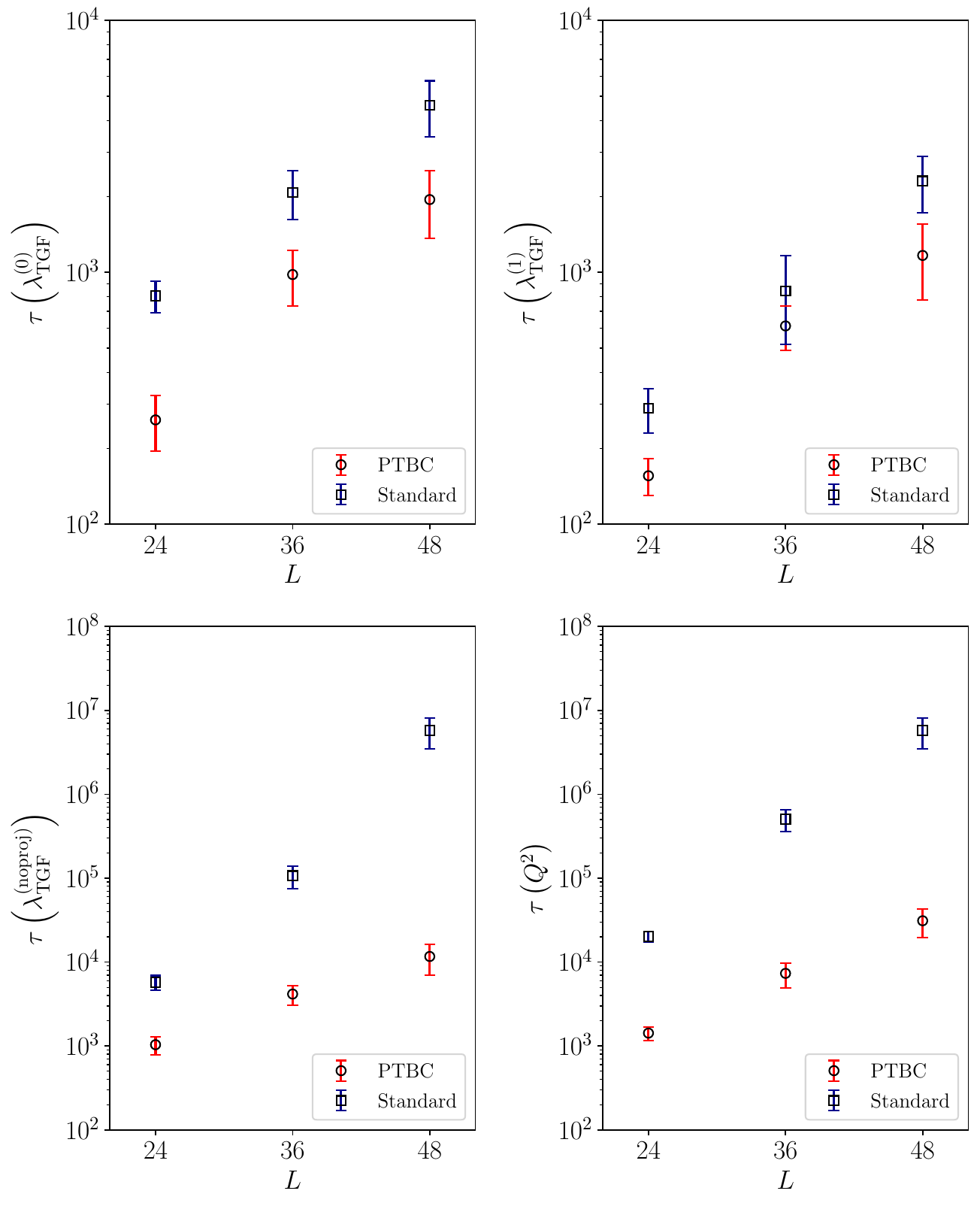}}
	\caption{Comparison of the integrated auto-correlation time of the projected and non-projected definitions of the couplings, and of the squared topological charge, obtained with the standard and the PTBC algorithms for the 3 simulations points corresponding to a LCP with $l=1.1$~fm. For the finest lattice spacing explored, only an upper bound on the auto-correlation time of $Q^2$ and of $\lambda_{\TGF}^{(\noproj)}$ could be set for the simulation with the standard algorithm, since no fluctuation of the topological charge was observed. In both cases, the auto-correlation time was expressed in units of lattice sweeps in order to make a fair comparison among the two algorithms. This means that the auto-correlation time was in both cases multiplied by the number of over-relaxation sweeps per heat-bath sweep, $n_{\ov}$, and, in the case of PTBC, also by the number of replicas $N_r$.}
	\label{fig:comp_tau_LCP2}
\end{figure*}

\begin{table*}[!t]
	\vspace*{0.5\baselineskip}
	\begin{center}
		\begin{tabular}{|c|c|c|c|c|c|}
			\hline
			\multicolumn{6}{|c|}{Parallel Tempering}\\
			\hline
			&&&&&\\[-1em]
			$L$ & $18 \times b$ & $\tau\left(\lambda_{\TGF}^{(0)}\right)$ & $\tau\left(\lambda_{\TGF}^{(1)}\right)$ & $\tau\left(\lambda_{\TGF}^{(\noproj)}\right)$ & $\tau\left(Q^2\right)$\\
			&&&&&\\[-1em]
			\hline
			24 & 6.459 & 260(65)   & 156(26)   & 1040(260)   & 1430(260)    \\
			36 & 6.765 & 980(250)  & 610(120)  & 4200(1100)  & 7350(2500)   \\ 
			48 & 6.992 & 1950(580) & 1170(390) & 11700(4700) & 31100(11700) \\
			\hline
		\end{tabular}
	\end{center}
	\begin{center}
		\begin{tabular}{|c|c|c|c|c|c|}
			\hline
			\multicolumn{6}{|c|}{Standard}\\
			\hline
			&&&&&\\[-1em]
			$L$ & $18 \times b$ & $\tau\left(\lambda_{\TGF}^{(0)}\right)$ & $\tau\left(\lambda_{\TGF}^{(1)}\right)$ & $\tau\left(\lambda_{\TGF}^{(\noproj)}\right)$ & $\tau\left(Q^2\right)$\\
			&&&&&\\[-1em]
			\hline
			24 & 6.459 & 800(120)   & 288(58)   & 5760(1150)            & 20200(2900)           \\
			36 & 6.765 & 2070(450)  & 840(320)  & 1.10(32)$\cdot 10^5$  & 5.1(1.5)$\cdot 10^5$ \\ 
			48 & 6.992 & 4600(1200) & 2300(580) & $\gtrsim 5.8(2.4) \cdot 10^6$  & $\gtrsim 5.8(2.4)\cdot 10^6$ \\
			\hline
		\end{tabular}
	\end{center}
		\caption{Integrated auto-correlation time of the projected and non-projected definitions of the couplings, and of the squared topological charge, obtained with the standard and the PTBC algorithm for the 3 simulations points corresponding to a LCP with $l=1.1$~fm. For the finest lattice spacing explored, only an upper bound on the auto-correlation time of $Q^2$ and of $\lambda_{\TGF}^{(\noproj)}$ could be set for the simulation with the standard algorithm, since no fluctuation of the topological charge was observed. In both cases, the auto-correlation time was expressed in units of lattice sweeps in order to make a fair comparison among the two algorithms. This means that the auto-correlation time was in both cases multiplied by the number of over-relaxation sweeps per heat-bath sweep, $n_{\ov}$, and, in the case of PTBC, also by the number of replicas $N_r$. Given that the obtained statistics are very large, and given that a precise assessment of the improvement of the PTBC algorithm has been already extensively discussed in previous works~\cite{Bonanno:2020hht,Bonanno:2022yjr,Bonanno:2023hhp,Bonanno:2024ggk}, here we simply relied on a standard binned jack-knife analysis to estimate $\tau$.}
	\label{tab:autocorr_times}
\end{table*}

The algorithmic improvement of PTBC can be quantified from the comparison of the auto-correlation times. Again, these are expressed in terms of lattice sweeps to account for the different number of over-relaxation sweeps per heat-bath sweep and the computational overhead introduced by the simulation of the unphysical replicas. Numerical results are reported in Tab.~\ref{tab:autocorr_times}, and shown in Fig.~\ref{fig:comp_tau_LCP2}. For what concerns the squared topological charge, we observe a reduction of the auto-correlation time $\tau$ by more than one order of magnitude for the coarsest lattice spacing and by more than two orders of magnitude for the finest one. Concerning the auto-correlation time of the non-projected coupling, we observe that the gain attained with PTBC is of the same order of magnitude, while for projected couplings is much smaller, as it is about a factor of $\sim 2-3$. Being the PTBC algorithm tailored to improve the evolution of the topological charge, this is a further indication, in addition to our results for the coupling, that the fluctuations of the global topological charge seem rather decoupled from those of the coupling once projected onto a fixed topological sector.

\subsection{Impact of topological freezing and topological projection on the step-scaling function}
\label{sec:res_LCP2}

We now aim at probing the impact of topological freezing and of topological projection on step-scaling following the strategy earlier outlined in Sec.~\ref{sec:strategy}, which will be now spelled out in more detail.

\begin{enumerate}
	
	\item We fix a target value for $\lambda_{\TGF}^{(0)}(\mu=2\mu_\had)=u^{(0)}_{\tg}$ on the LCP1. Since we have shown that the projected and non-projected couplings give consistent results using both algorithms, we can choose any of the determinations of the previous section to fix $u^{(0)}_{\tg}$. We chose:
	\be
	u^{(0)}_\tg \equiv \lambda_{\TGF}^{(0)}(18b=7,L=24)\bigg\vert_{\rm Standard} = 13.9063406.
	\ee

	\item We now consider the results for the renormalized coupling obtained for the same bare couplings of LCP1, but on doubled lattices. Our goal is to compute the continuum step-scaling functions corresponding to projected and non-projected couplings at $\mu_\had$,
	\beq
	\sigma_{\TGF}^{(0)} (u^{(0)}_\tg)  \text{ and } \sigma_{\TGF}^{(\noproj)} (u^{(0)}_\tg)\,.
	\eeq
	This is done following the same procedure put forward in Ref.~\cite{Bribian:2021cmg}, spelled out here for clarity. Since the tuning of the lattices is not perfect, there is a small mismatch in the lattice determined values of $u^{(0)}_{\tg}$. To correct for that, we slightly shift the values of the lattice step-scaling function taking into account the shifts in $u$ required to match $u_\tg$. This is done according to the formula:
	\beq\label{eq:sigma_tg}
	\begin{aligned}
		\Sigma\left(u^{(0)}_\tg,L\right) =  \Sigma\left(u,L\right) - \frac{\Sigma^2\left(u,L\right)}{u^2}\left(u-u^{(0)}_{\tg}\right),
	\end{aligned}
	\eeq
	where $\Sigma(u,L)$ stands for the value of the coupling obtained for the simulation point $(b,2L)$ and corresponding to the coupling $u$ obtained for the simulation point $(b,L)$. The relation used to determine $\Sigma\left(u^{(0)}_\tg,L\right)$ follows from the fact that, at leading order of perturbation theory, one expects $1/\Sigma(u) - 1/u = \mathrm{constant}$.
	Finally, the values of $\Sigma(u^{(0)}_\tg,L)$ are extrapolated to the continuum limit, defining
	\beq\label{eq:sigma_tg_def}
	\sigma(u^{(0)}_{\tg}) = \lim_{1/L \, \to \, 0} \Sigma(u^{(0)}_{\tg},L)\,.
	\eeq
	In Fig.~\ref{fig:sigma_target_vol2} we report the continuum extrapolations of $\Sigma(u^{(0)}_{\tg},L)$ obtained using the data shown in Sec.~\ref{sec:res_LCP1} for the $Q=0$ projected couplings, and obtained from the two different algorithms. The results are perfectly consistent:
	\beq
	\label{eq:sigma_tg_Q0_std}
	\sigma_{\TGF}^{(0)} (u^{(0)}_\tg) &= 34.43(24) \qquad &\text{(Standard)},\\
	\sigma_{\TGF}^{(0)} (u^{(0)}_\tg) &= 34.61(29) \qquad &\text{(PTBC)}\, .
	\eeq
	Repeating the procedure with the non-projected coupling, which could only be computed reliably using the PTBC algorithm, the continuum extrapolated step-scaling function we obtain is:
	\beq\label{eq:sigma_tg_noproj_PTBC}
	\sigma_{\TGF}^{(\noproj)} (u^{(0)}_\tg)=36.31(26) \qquad &\text{(PTBC)}\, .
	\eeq
	
	\begin{figure}[!t]
		\centering
		\resizebox{0.45\textwidth}{!}{\includegraphics{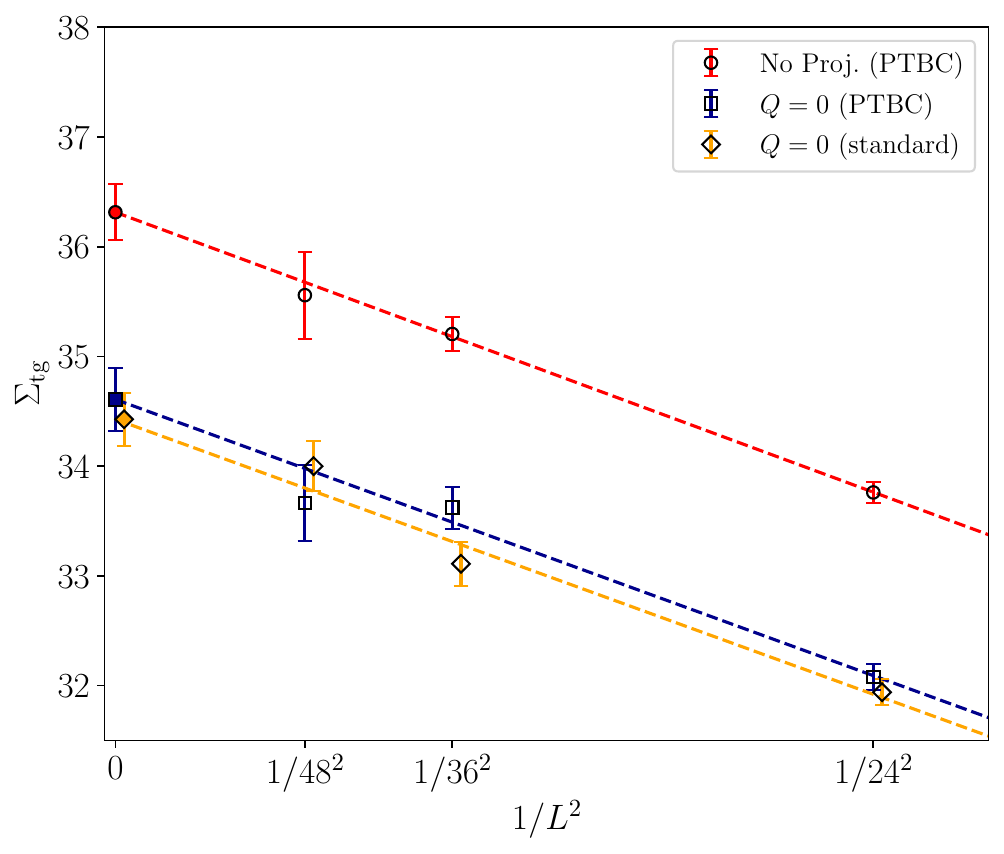}}
		\caption{Continuum limit extrapolation of $\Sigma_\tg \equiv \Sigma(u^{(0)}_\tg,L)$ calculated using Eq.~\eqref{eq:sigma_tg_def} from a projected and an non-projected definition of the coupling. In the former case, we show results obtained both using the PTBC and the standard algorithms.}
		\label{fig:sigma_target_vol2}
	\end{figure}
	
	\item As already discussed, topological fluctuations become relevant with this lattice volume, leading to different projected and non-projected couplings. Therefore, at this point, topological freezing and topological projection could have a stronger effect on the step-scaling sequence. We now want to check whether, under the step-scaling sequence realized with the previous steps, the renormalization scale changes consistently for the projected and the non-projected couplings.
	
	With this purpose in mind, we determined the bare couplings $b$ that lead to a value of the projected coupling equal to $\sigma_{\TGF}^{(0)} (u^{(0)}_\tg)=34.43$, cf.~Eq.~\eqref{eq:sigma_tg_Q0_std}.~\footnote{The tuning of the bare couplings $b$ is done as indicated in Sec.~3.4 of Ref.~\cite{Bribian:2021cmg}. It is based on fitting, for each $L$, the dependence on $b$ of the projected couplings obtained using the standard algorithm.} This defines the line of constant physics dubbed as LCP2 in Tab.~\ref{tab:summary_simulations}. If the projection of the coupling does not introduce a bias, and thus it is a legitimate and consistent way of defining an LCP, this should also be a proper LCP for the non-projected coupling, leading to a value that satisfies Eq.~\eqref{eq:test} with $\sigma_{\TGF}^{(\noproj)} (u^{(0)}_\tg)=36.31(26)$, cf.~Eq.~\eqref{eq:sigma_tg_noproj_PTBC}.
	
\end{enumerate}

\begin{figure*}[!t]
	\centering
	\resizebox{0.65\textwidth}{!}{\includegraphics{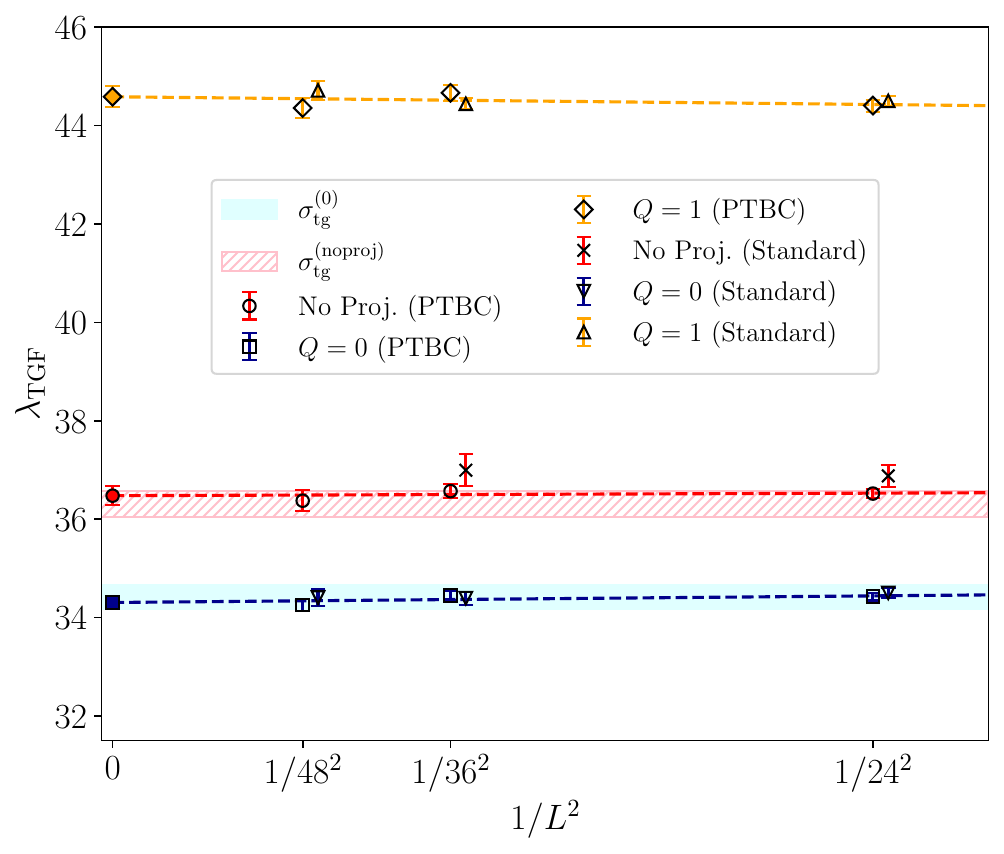}}
	\caption{Extrapolation towards the continuum limit of the projected and non-projected couplings obtained with the PTBC algorithm for the 3 simulation points corresponding to a LCP with $l\simeq 1.1$~fm, tuned to achieve a constant value of $\sigma_{\TGF}^{(0)} (u^{(0)}_\tg) = 34.43(24)$ (uniform shaded band), compared with the results obtained with the standard algorithm. The dashed shaded area represents $\sigma_{\TGF}^{(\noproj)} (u^{(0)}_\tg)=36.31(26)$ obtained with the PTBC algorithm, cf.~Eq.~\eqref{eq:sigma_tg_noproj_PTBC}.}
	\label{fig:LCP2_tuned_coupling}
\end{figure*}

Finally, let us present the results of this test. The determination of the projected and non-projected couplings obtained on the LCP2 is shown in Fig.~\ref{fig:LCP2_tuned_coupling}. The results obtained with parallel tempering for the non-projected coupling show no visible dependence on $L$ within the achieved per mil accuracy, and agree perfectly with the continuum-extrapolated target value $\sigma_{\TGF}^{(\noproj)} (u^{(0)}_\tg)=36.31(26)$ earlier obtained in Eq.~\eqref{eq:sigma_tg_noproj_PTBC}, represented by the dashed shaded area in the plot. Moreover, also in this case we find perfect agreement between the results obtained with the PTBC and the standard algorithms for the projected couplings. This piece of evidence completes the plan outlined at the beginning of this section, and fully confirms the reliability of topological projection for the purpose of calculating the step-scaling function and hence the $\Lambda$-parameter.

\subsection{Behavior of topological quantities along the LCP}

\begin{figure*}[!t]
	\centering
	\resizebox{0.45\textwidth}{!}{\includegraphics{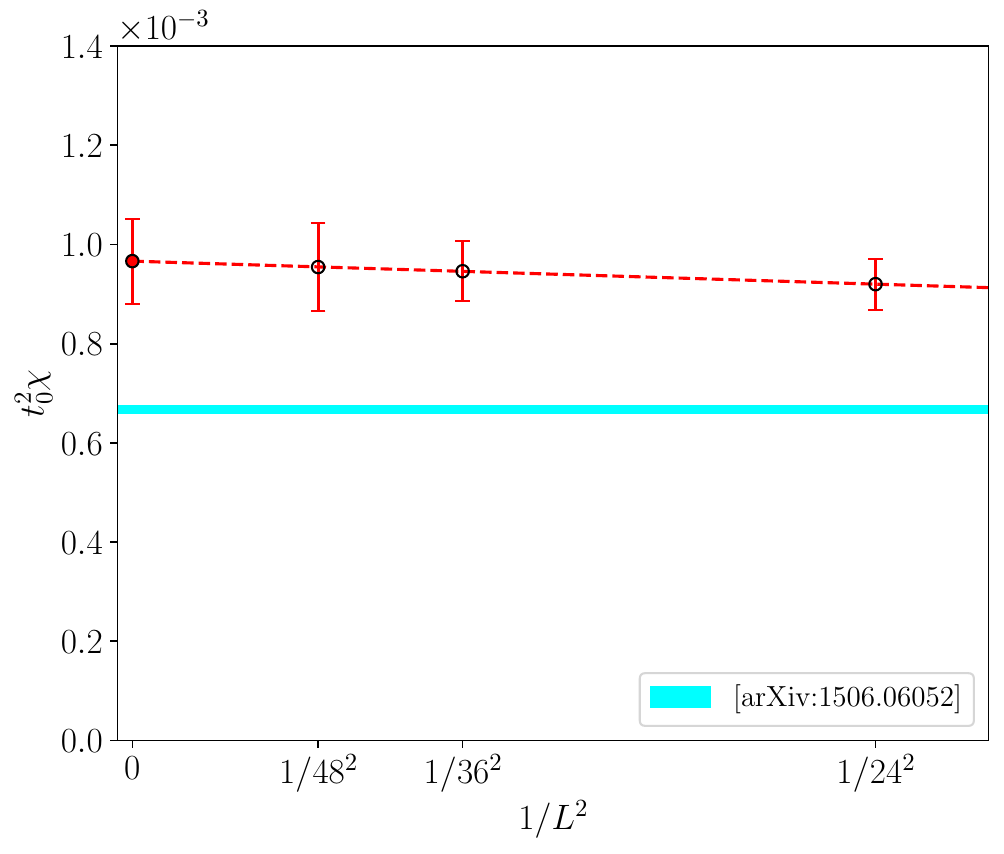}}
	\resizebox{0.45\textwidth}{!}{\includegraphics{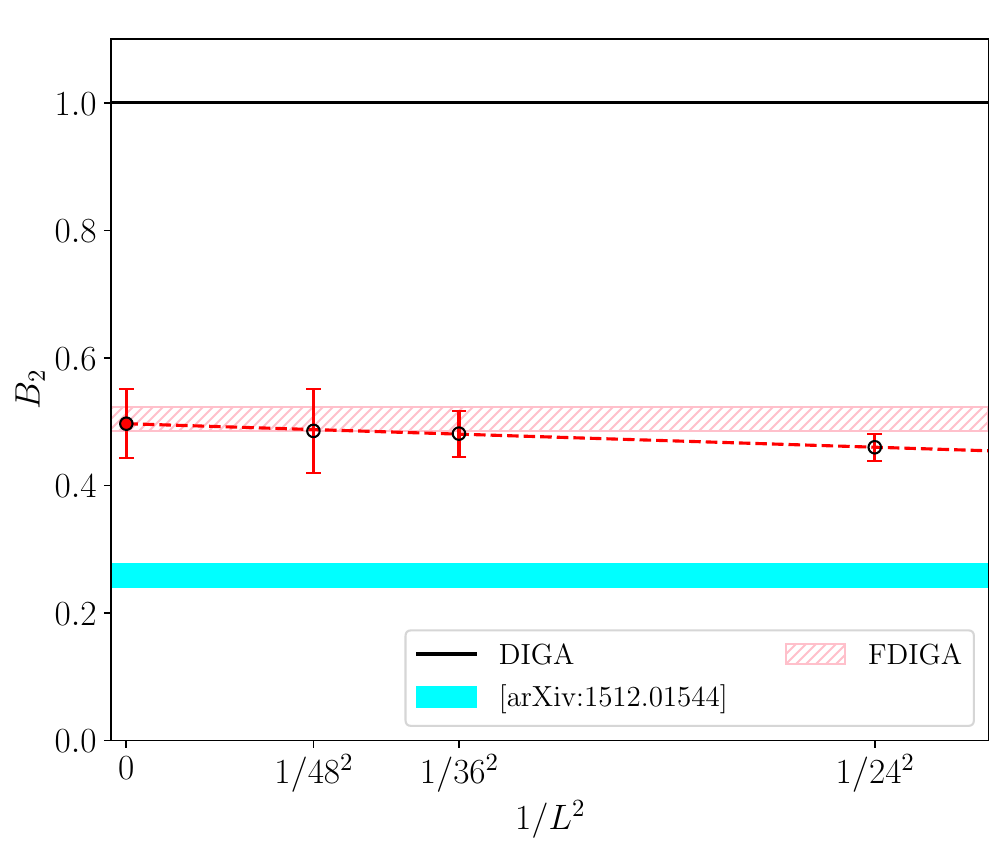}}
	\caption{Continuum limit of the topological susceptibility $t_0^2 \chi$ and of the quartic coefficient $B_2$ obtained with the PTBC algorithm for the 3 simulations points corresponding to a LCP with $l\simeq 1.1$~fm. The infinite-volume estimates of $t_0^2 \chi$ and of $B_2$, displayed as uniform shaded areas, are taken from Refs.~\cite{Ce:2015qha,Bonati:2015sqt}. The dashed shaded area and the solid line represent instead the predictions for $B_2$ obtained using, respectively, the FDIGA and the DIGA, see the text.}
	\label{fig:LCP2_tuned_topo}
\end{figure*}

As a by-product of our investigation, by virtue of the adoption of the parallel tempering algorithm, we were also able to reliably compute two topological observables, namely, the topological susceptibility,
\beq
t_0^2 \chi = \left(\frac{\sqrt{t_0}}{a}\right)^4 \frac{\braket{Q^2}}{\tL^2 L^2},
\eeq
and the dimensionless quartic coefficient $B_2$,\footnote{Sometimes, the different definition $b_2 = - B_2/12$ is employed in the literature.}
\beq
B_2 = \frac{\braket{Q^4}-3\braket{Q^2}^2}{\braket{Q^2}}.
\eeq

These two quantities are tightly related to the dependence of the vacuum energy $E$ on the dimensionless parameter $\theta$, coupling the topological charge to the pure-gauge action,
\beqnn
E(\theta) &\equiv& -\frac{1}{V}\log\left[\frac{Z(\theta)}{Z(0)}\right],\\ Z(\theta) &\equiv& \int [dA] e^{-S_{\YM}[A] + i \theta Q[A]},\\
E(\theta) &=& \frac{1}{2}\chi \theta^2\left[1-\frac{1}{12}B_2\theta^2 + \mathcal{O}\left(\theta^4\right)\right],
\eeqnn
and have several important theoretical and phenomenological implications~\cite{Ce:2015qha,Borsanyi:2015cka,Petreczky:2016vrs,Athenodorou:2022aay,Bonati:2015sqt,Witten:1979vv,Veneziano:1979ec,Alles:1996nm,Alles:1997qe,DelDebbio:2002xa,DelDebbio:2004ns,DElia:2003zne,Lucini:2004yh,Giusti:2007tu,Vicari:2008jw,Panagopoulos:2011rb,Bonati:2013tt,Ce:2016awn,Berkowitz:2015aua,Bonati:2016tvi,Bonati:2018rfg,Bonati:2019kmf}.

First of all, we observe that both quantities show extremely mild lattice artefacts when computed along the LCP2, as it can be seen from Fig.~\ref{fig:LCP2_tuned_topo}. This is yet a further confirmation that the calibration of the LCP done according to the $Q=0$ projected coupling is a legitimate LCP also for topology-related quantities. This is actually not surprising, as, of course, $\braket{Q^2}$ and $\lambda_{\TGF}^{(\noproj)}$ are not unrelated. In particular, recalling that we are working in a regime where $\braket{Q^2}$ is small, the following approximate relation holds:
\beq
\begin{aligned}
	\lambda_{\TGF}^{(\noproj)} &\simeq \frac{P_0 \lambda_{\TGF}^{(0)} + 2 P_1 \lambda_{\TGF}^{(1)}}{P_0+2P_1} \\
	&\simeq \frac{1}{1+\braket{Q^2}}\lambda_{\TGF}^{(0)} + \braket{Q^2} \lambda_{\TGF}^{(1)}, \qquad \\
	\braket{Q^2} &\simeq 2\frac{P_1}{P_0}.
\end{aligned}
\eeq

We also would like to remark that our lattice results for $\chi$ and $B_2$ cannot be directly compared with previous determinations obtained in the literature. As a matter of fact, here we are working in an intermediate semi-classical regime with a volume $l \sim 1.1$ fm, and significant finite-size effects are expected for topological quantities for lattice sizes below $\sim 1.4-1.5$ fm.

Indeed, we observe that our continuum determination of the topological susceptibility $t_0^2\chi=9.66(88)\cdot 10^{-4}$ differs from the one obtained in Ref.~\cite{Ce:2015qha} on a much larger volume, $t_0^2\chi=6.67(7)\cdot 10^{-4}$, see Fig.~\ref{fig:LCP2_tuned_topo} on the left. Finite-volume effects can be also seen in our continuum determination of $B_2=0.497(54)$, which also differs from the large-volume result of Ref.~\cite{Bonati:2015sqt}, $B_2=0.259(30)$, see Fig.~\ref{fig:LCP2_tuned_topo} on the right.

However, one can use semi-classical methods to obtain a prediction for $B_2$ based on the free instanton gas. Indeed, by virtue of the volume-suppression of topological 
fluctuations\footnote{It is interesting to notice that $\langle Q^2 \rangle$ is a decreasing function of the volume $V$, despite the 
	fact that the topological susceptibility $\chi = \langle Q^2 \rangle / V$ shows instead a peak in 
	the semi-classical regime (see, e.g., Refs.~\cite{Bribian:2021cmg} and~\cite{Gonzalez-Arroyo:1995ynx}): indeed, as stated above, our continuum determination of 
	$\chi$ is larger than that obtained in the large-volume limit.}, in this regime one can describe the Yang--Mills vacuum in terms of a dilute gas of weakly-interacting quasi-particles possessing a non-trivial topological charge.

More precisely, there is plenty of theoretical and numerical evidence pointing out that, in the presence of TBCs, the semi-classical regime of $\SU(N)$ Yang--Mills theories can be accurately described in terms of \emph{fractional instantons}, i.e., topological objects with $Q=\pm 1/N$, see Ref.~\cite{Gonzalez-Arroyo:2023kqv} for a recent review and for further references.

The details of the calculation of the $\theta$-dependent vacuum energy from the so-called Fractional Dilute Instanton Gas Approximation (FDIGA) can be found in App.~\ref{app:FDIGA}. Here, we just limit to say that $E(\theta)$ at the semi-classical level depends only on one parameter: $\braket{Q^2} = V \chi$. Thus, fixing it from our lattice result for $\chi$ we are able to obtain a semi-classical FDIGA prediction for $B_2$.

This procedure yields, for the particular value of $\langle Q^2 \rangle$ obtained in our simulations, $B_2^{(\rm FDIGA)}=0.504(19)$, which is in excellent agreement with our lattice determination, cf.~the left panel of Fig.~\ref{fig:LCP2_tuned_topo}. On the other hand, the ordinary Dilute Instanton Gas Approximation (DIGA)~\cite{Gross:1980br,Schafer:1996wv} (based on $Q=\pm 1$ charged topological objects) does not clearly work in this regime, as it would yield $B_2^{(\rm DIGA)}=1$.

\section{Conclusions}\label{sec:conclu}

We have presented a new investigation of the role played by topology in the determination of the renormalized strong coupling constant from lattice simulations, with the goal of assessing the possible systematic effects introduced by topological freezing and topological projection in the determination of the $\SU(3)$ pure-Yang--Mills $\Lambda$-parameter.

Our investigation combines twisted volume reduction and the gradient flow according to the setup of Ref.~\cite{Bribian:2021cmg}. We employed the $\SU(N)$ Parallel Tempering on Boundary Conditions algorithm of Ref.~\cite{Bonanno:2020hht}, suitably generalized to include TBCs, to accurately determine the step-scaling function, corresponding to the sequence $\mu_{\had} \to 2 \mu_{\had}$, avoiding the effects of topological freezing. As a matter of fact, the PTBC algorithm allows to achieve a reduction of the auto-correlation time of the topological charge by up to two orders of magnitude compared to the standard algorithm.

Results obtained with parallel tempering show that the topological projection works for the strong coupling even in the presence of severe topological freezing. Indeed, results obtained for the $Q=0$ projected coupling with both algorithms always turn out to be in perfect agreement among themselves at the per mil accuracy we reached. Moreover, we showed that an LCP defined to have a fixed projected renormalized coupling is also an LCP for the non-projected renormalized coupling. Other topology-related quantities such as the topological susceptibility, which only the PTBC algorithm allowed to determine, also share the same LCP. These findings imply that topological projection in the presence of topological freezing leads to the same step-scaling sequence that would have been otherwise obtained with the non-projected coupling, and thus ultimately to the same $\Lambda$-parameter.

Our current results can be expanded in several directions. For instance, it would be interesting to further investigate topological projection on small volumes by combining our twisted PTBC setup with the multicanonical algorithm, which can largely improve the sampling of rare volume-suppressed topological fluctuations. It would also be very interesting to extend our investigation to larger values of $N$ in order to study the large-$N$ limit of the $\Lambda$-parameter, which is an extremely interesting theoretical topic. To this end, it is crucial to check the effects of topological freezing on the scale setting procedure at large $N$, which can be efficiently achieved adopting the PTBC algorithm.

\section*{Acknowledgements}
It is a pleasure to thank Alberto Ramos for useful discussions and comments, and also for carefully reading this manuscript. This work is partially supported by the Spanish Research Agency (Agencia Estatal de Investigación) through the grant IFT Centro de Excelencia Severo Ochoa CEX2020-001007-S and, partially, by the grant PID2021-127526NB-I00, both of which are funded by MCIN/AEI/10.13039/501100011033. This work has also been partially supported by the project ”Non-perturbative aspects of fundamental interactions, in the Standard Model and beyond” funded by MUR, Progetti di Ricerca di Rilevante Interesse Nazionale (PRIN), Bando 2022, grant 2022TJFCYB (CUP I53D23001440006). We also acknowledge partial support from the project H2020-MSCAITN-2018-813942 (EuroPLEx) and the EU Horizon 2020 research and innovation programme, STRONG-2020 project, under grant agreement No. 824093. Numerical calculations have been performed partially on the \texttt{Marconi} machine at Cineca, based on the agreement between INFN and Cineca, under the projects INF22\_npqcd and INF23\_npqcd, and partially on the \texttt{Finisterrae~III} cluster at CESGA (Centro de Supercomputaci\'on de Galicia).

\appendix

\begin{table*}[!t]
	\footnotesize
	\begin{center}
		\begin{tabular}{|c|c|c|c|c|c|}
			\hline
			\multicolumn{6}{|c|}{Parallel Tempering}\\
			\hline
		\end{tabular}
	\end{center}
	\begin{center}
		\begin{tabular}{|c|c|c|c|}
			\hline
			\multicolumn{4}{|c|}{}\\[-1em]
			\multicolumn{4}{|c|}{LCP1 ($l=0.55$~fm)}\\
			\hline
			&&&\\[-1em]
			$L$ & $18 \times b$ & $\lambda_{\TGF}^{(\noproj)}$ & $\lambda_{\TGF}^{(0)}$ \\
			&&&\\[-1em]
			\hline
			12 & 6.4881 & 13.9679(44) & 13.9503(47) \\
			18 & 6.7790 & 13.945(18)  & 13.936(20)  \\ 
			24 & 7.0000 & 13.945(25)  & 13.953(29)  \\
			\hline
		\end{tabular}
	\end{center}
	\begin{center}
		\begin{tabular}{|c|c|c|c|c|c|}
			\hline
			\multicolumn{5}{|c|}{}\\[-1em]
			\multicolumn{5}{|c|}{LCP1 with doubled $L$}\\
			\hline
			&&&&\\[-1em]
			$L$ & $18 \times b$ & $\lambda_{\TGF}^{(\noproj)}$ & $\lambda_{\TGF}^{(0)}$ & $\lambda_{\TGF}^{(1)}$ \\
			&&&&\\[-1em]
			\hline
			24 & 6.459 & 34.128(90) & 32.31(12) & 42.78(18) \\
			36 & 6.765 & 35.46(11)  & 33.80(15) & 43.10(23) \\ 
			48 & 6.992 & 35.81(36)  & 33.94(30) & 43.35(18) \\
			\hline
		\end{tabular}
	\end{center}
	\begin{center}
		\begin{tabular}{|c|c|c|c|c|c|c|}
			\hline
			\multicolumn{7}{|c|}{}\\[-1em]
			\multicolumn{7}{|c|}{LCP2 ($l=1.1$~fm)}\\
			\hline
			&&&&&&\\[-1em]
			$L$ & $18 \times b$ & $\lambda_{\TGF}^{(\noproj)}$ & $\lambda_{\TGF}^{(0)}$ & $\lambda_{\TGF}^{(1)}$ & $\braket{Q^2}$ & $B_2$\\
			&&&&\\[-1em]
			\hline
			24 & 6.459 & 36.520(96) & 34.426(81) & 44.41(12) & 0.1997(46) & 0.460(21) \\
			36 & 6.765 & 36.57(14)  & 34.45(10)  & 44.67(16) & 0.2050(78) & 0.481(36) \\ 
			48 & 6.992 & 36.37(21)  & 34.26(12)  & 44.36(21) & 0.207(16)  & 0.485(66) \\
			\hline
		\end{tabular}
	\end{center}
	\caption{Summary of the obtained results using the PTBC algorithm.}
	\label{tab:raw_data_PTBC}
\end{table*}

\begin{table*}[!t]
	\footnotesize
	\begin{center}
		\begin{tabular}{|c|c|c|c|c|c|}
			\hline
			\multicolumn{6}{|c|}{Standard}\\
			\hline
		\end{tabular}
	\end{center}
	\begin{center}
		\begin{tabular}{|c|c|c|c|}
			\hline
			\multicolumn{4}{|c|}{}\\[-1em]
			\multicolumn{4}{|c|}{LCP1 ($l=0.55$~fm)}\\
			\hline
			&&&\\[-1em]
			$L$ & $18 \times b$ & $\lambda_{\TGF}^{(\noproj)}$ & $\lambda_{\TGF}^{(0)}$ \\
			&&&\\[-1em]
			\hline
			12 & 6.4881 & 13.971(13) & 13.948(11) \\
			18 & 6.7790 & 13.939(13) & 13.938(11) \\ 
			24 & 7.0000 & 13.903(26) & 13.906(22) \\
			\hline
		\end{tabular}
	\end{center}
	\begin{center}
		\begin{tabular}{|c|c|c|c|c|c|c|c|}
			\hline
			\multicolumn{5}{|c|}{}\\[-1em]
			\multicolumn{5}{|c|}{LCP1 with doubled $L$}\\
			\hline
			&&&&\\[-1em]
			$L$ & $18 \times b$ & $\lambda_{\TGF}^{(\noproj)}$ & $\lambda_{\TGF}^{(0)}$ & $\lambda_{\TGF}^{(1)}$ \\
			&&&&\\[-1em]
			\hline
			24 & 6.4881 & 33.94(21) & 32.17(11) & 42.96(23) \\
			36 & 6.7990 & 35.19(99) & 33.29(19) & 43.85(41) \\ 
			48 & 7.0000 &     -     & 34.00(20) & 43.94(90) \\
			\hline
		\end{tabular}
	\end{center}
	\begin{center}
		\begin{tabular}{|c|c|c|c|c|}
			\hline
			\multicolumn{5}{|c|}{}\\[-1em]
			\multicolumn{5}{|c|}{LCP2 ($l=1.1$~fm)}\\
			\hline
			&&&&\\[-1em]
			$L$ & $18 \times b$ & $\lambda_{\TGF}^{(\noproj)}$ & $\lambda_{\TGF}^{(0)}$ & $\lambda_{\TGF}^{(1)}$ \\
			&&&&\\[-1em]
			\hline
			24 & 6.459 & 36.88(23) & 34.501(98) & 44.50(11) \\
			36 & 6.765 & 37.00(33) & 34.39(13)  & 44.44(11) \\ 
			48 & 6.992 & -         & 34.31(17)  & 44.72(20) \\
			\hline
		\end{tabular}
	\end{center}
	\caption{Summary of the obtained results using the standard algorithm.}
	\label{tab:raw_data_STD}
\end{table*}

\section*{Appendix}

\section{Raw data}\label{app:raw_data}

In this appendix we report in Tabs.~\ref{tab:raw_data_PTBC} and~\ref{tab:raw_data_STD} all the numerical results shown in the plots in the main text. These two tables refer, respectively, to the PTBC and to the standard algorithms.

\section{Dependence on $\theta$ with Fractional Dilute Instanton Gas Approximation}\label{app:FDIGA}

In this appendix we show how to derive the semiclassical expressions for the topological susceptibility and the quartic coefficient $B_2$ in the Fractional Dilute Instanton Gas Approximation (FDIGA). The main difference with the standard DIGA is the fact that the dilute gas is composed of instantons with fractional charge $Q=\pm 1/N = \pm 1/3$, as opposed to ordinary $Q=\pm 1$ instantons. These objects arise in a natural way on a torus with twisted boundary conditions and non-orthogonal twist ($n_{\mu \nu} \tilde n_{\mu \nu} \neq 0 \, (\text{mod } N)$)~\cite{tHooft:1979rtg} and have been the basis of the instanton liquid model of confinement put forward by Gonz\'alez-Arroyo and collaborators; for a recent review and further references see~\cite{Gonzalez-Arroyo:2023kqv}, see also~\cite{vanBaal:1984ra,RTN:1993ilw,GarciaPerez:1993jw,Gonzalez-Arroyo:1995ynx,vanBaal:2000zc,Unsal:2020yeh,Cox:2021vsa,Bribian:2021cmg,Nair:2022yqi}. Although our setup corresponds to an orthogonal twist, fractional instantons may still arise, provided that their total contribution to the topological charge amounts to an integer, and they likely represent a good description of the topological activity of the medium in terms of weakly-interacting objects. Indeed, in this context, Ref.~\cite{Bribian:2021cmg} showed how the correlations observed in their small-to-intermediate volume TBC simulations between topological charge and coupling were quantitatively well described in this approximation. In this work, we have extended the analysis to the determination of the $B_2$ coefficient, showing that the prediction provided by FDIGA also works very well for this quantity in the appropriate regime. 

The starting point for extracting the desired quantities is the dilute gas approximation for $\SU(N)$ fractional instantons.  As we have chosen an orthogonal twist, the total topological charge remains quantized in integer units. This constraint must be applied when formulating the dilute gas fractional instanton partition function, which, when restricted to the sector of topological charge $Q$, reads as follows:
\beq
Z_Q = {\cal C}  \sum_{n,\overline{n}} \frac{1}{n! \overline{n}!}  ( RV)^{n +\overline{n}} \delta(n-\overline{n} -N Q)
\, ,
\eeq
where $R$ stands for the probability of creating a fractional instanton per unit volume $V$. From this expression, the 
$\theta$-dependent FDIGA partition function can be easily derived to be~\cite{vanBaal:1984ra,vanBaal:2000zc,Bribian:2021cmg}:
\be
Z(\theta) \equiv \sum_{Q\in \mathbf{Z}}  e^{i Q \theta}  Z_Q = \frac{{\cal  C }}{N} \sum_{k=1}^{N} \exp\left \{ x \cos \left (\frac{\theta+ 2 \pi k}{N}\right ) \right \}
\, ,
\ee
where $x=2RV$; the reader is referred to Ref.~\cite{Bribian:2021cmg} for more details on how to derive this expression. Taking derivatives of $Z(\theta)=Z(0) \, e^{-VE(\theta)}$ with respect to $\theta$, it is now trivial to derive the expressions for $\langle Q^2\rangle$ and $\langle Q^4\rangle$ in this approximation. For the  $\SU(3)$ topological susceptibility, one obtains for instance:
\beq\label{eq:chi_FDIGA}
V\chi(x) = \braket{Q^2}(x) = \frac{x}{18} \, \left ( 2 - \frac{3 (2 + x)}{2 +  e^{3 x/2}} \right ) \, ,
\eeq
while the result for $B_2$ in $\SU(3)$ is given by:
\beq\label{eq:B2_FDIGA}
B_2(x) = \frac{1}{9} + \frac{x (2 + x)} {2 (2 + e^{3 x/2})} - \frac{x^2 (8 + x)}
{ 8 - 8 e^{3 x/2} + 12 x}.
\eeq
It is interesting to consider two relevant limits of Eq.~(\ref{eq:B2_FDIGA}). 
For $x \to \infty$, which corresponds to the large-volume limit, 
one obtains $B_2 = 1/9$, which is the value expected for an unconstrained 
FDIGA, i.e., a gas of non-interacting objects of charge $\pm 1/3$: indeed, in this limit the
overall constraint of integer $Q$ becomes irrelevant. In the opposite limit of small $x$ 
(small-volume limit) one obtains $B_2 = 1$, which corresponds to the DIGA prediction and is reproduced
fictitiously in this limit, just because $Q =0$ most of the time with only very rare fluctuations
to $Q = \pm 1$, which, as a matter of fact, is Poissonian (hence reproduces DIGA).

To determine $B_2$ from Eq.~\eqref{eq:B2_FDIGA} in intermediate cases like ours, it is necessary to first determine the input quantity $x=2RV$. Given the unreliability of the semi-classical approximation to this end, this can be done by inverting Eq.~\eqref{eq:chi_FDIGA}, using the value of the topological susceptibility measured on the lattice as input. More precisely, for the lattices corresponding to the LCP2, we found $\braket{Q^2} = 0.209(12)$, leading to $x=2.322(82)$ and to $B_2=0.504(19)$.

\providecommand{\href}[2]{#2}\begingroup\raggedright\endgroup


\begin{thebibliography}{100}
	
	\bibitem{DallaBrida:2020pag}
	M.~Dalla~Brida, \emph{{Past, present, and future of precision determinations of
			the QCD parameters from lattice QCD}},
	\href{https://doi.org/10.1140/epja/s10050-021-00381-3}{\emph{Eur. Phys. J. A}
		{\bfseries 57} (2021) 66} [\href{https://arxiv.org/abs/2012.01232}{{\ttfamily
			2012.01232}}].
	
	\bibitem{Maltman:2008bx}
	K.~Maltman, D.~Leinweber, P.~Moran and A.~Sternbeck, \emph{{The Realistic
			Lattice Determination of $\alpha(s)(M(Z))$ Revisited}},
	\href{https://doi.org/10.1103/PhysRevD.78.114504}{\emph{Phys. Rev. D}
		{\bfseries 78} (2008) 114504}
	[\href{https://arxiv.org/abs/0807.2020}{{\ttfamily 0807.2020}}].
	
	\bibitem{PACS-CS:2009zxm}
	{\scshape PACS-CS} collaboration, S.~Aoki et~al., \emph{{Precise determination
			of the strong coupling constant in $N_f$ = 2+1 lattice QCD with the
			Schrodinger functional scheme}},
	\href{https://doi.org/10.1088/1126-6708/2009/10/053}{\emph{JHEP} {\bfseries
			10} (2009) 053} [\href{https://arxiv.org/abs/0906.3906}{{\ttfamily
			0906.3906}}].
	
	\bibitem{McNeile:2010ji}
	C.~McNeile, C.~T.~H. Davies, E.~Follana, K.~Hornbostel and G.~P. Lepage,
	\emph{{High-Precision c and b Masses, and QCD Coupling from Current-Current
			Correlators in Lattice and Continuum QCD}},
	\href{https://doi.org/10.1103/PhysRevD.82.034512}{\emph{Phys. Rev. D}
		{\bfseries 82} (2010) 034512}
	[\href{https://arxiv.org/abs/1004.4285}{{\ttfamily 1004.4285}}].
	
	\bibitem{Chakraborty:2014aca}
	B.~Chakraborty, C.~T.~H. Davies, B.~Galloway, P.~Knecht, J.~Koponen, G.~C.
	Donald et~al., \emph{{High-precision quark masses and QCD coupling from
			$n_f=4$ lattice QCD}},
	\href{https://doi.org/10.1103/PhysRevD.91.054508}{\emph{Phys. Rev. D}
		{\bfseries 91} (2015) 054508}
	[\href{https://arxiv.org/abs/1408.4169}{{\ttfamily 1408.4169}}].
	
	\bibitem{Bruno:2017gxd}
	{\scshape ALPHA} collaboration, M.~Bruno, M.~Dalla~Brida, P.~Fritzsch,
	T.~Korzec, A.~Ramos, S.~Schaefer et~al., \emph{{QCD Coupling from a
			Nonperturbative Determination of the Three-Flavor $\Lambda$ Parameter}},
	\href{https://doi.org/10.1103/PhysRevLett.119.102001}{\emph{Phys. Rev. Lett.}
		{\bfseries 119} (2017) 102001}
	[\href{https://arxiv.org/abs/1706.03821}{{\ttfamily 1706.03821}}].
	
	\bibitem{Cali:2020hrj}
	S.~Cali, K.~Cichy, P.~Korcyl and J.~Simeth, \emph{{Running coupling constant
			from position-space current-current correlation functions in three-flavor
			lattice QCD}},
	\href{https://doi.org/10.1103/PhysRevLett.125.242002}{\emph{Phys. Rev. Lett.}
		{\bfseries 125} (2020) 242002}
	[\href{https://arxiv.org/abs/2003.05781}{{\ttfamily 2003.05781}}].
	
	\bibitem{Bazavov:2019qoo}
	{\scshape TUMQCD} collaboration, A.~Bazavov, N.~Brambilla, X.~Garcia~i Tormo,
	P.~Petreczky, J.~Soto, A.~Vairo et~al., \emph{{Determination of the QCD
			coupling from the static energy and the free energy}},
	\href{https://doi.org/10.1103/PhysRevD.100.114511}{\emph{Phys. Rev. D}
		{\bfseries 100} (2019) 114511}
	[\href{https://arxiv.org/abs/1907.11747}{{\ttfamily 1907.11747}}].
	
	\bibitem{Ayala:2020odx}
	C.~Ayala, X.~Lobregat and A.~Pineda, \emph{{Determination of $\alpha(M_z)$ from
			an hyperasymptotic approximation to the energy of a static quark-antiquark
			pair}}, \href{https://doi.org/10.1007/JHEP09(2020)016}{\emph{JHEP} {\bfseries
			09} (2020) 016} [\href{https://arxiv.org/abs/2005.12301}{{\ttfamily
			2005.12301}}].
	
	\bibitem{FlavourLatticeAveragingGroupFLAG:2021npn}
	{\scshape Flavour Lattice Averaging Group (FLAG)} collaboration, Y.~Aoki
	et~al., \emph{{FLAG Review 2021}},
	\href{https://doi.org/10.1140/epjc/s10052-022-10536-1}{\emph{Eur. Phys. J. C}
		{\bfseries 82} (2022) 869}
	[\href{https://arxiv.org/abs/2111.09849}{{\ttfamily 2111.09849}}].
	
	\bibitem{Workman:2022ynf}
	{\scshape Particle Data Group} collaboration, R.~L. Workman and Others,
	\emph{{Review of Particle Physics}},
	\href{https://doi.org/10.1093/ptep/ptac097}{\emph{PTEP} {\bfseries 2022}
		(2022) 083C01}.
	
	\bibitem{DallaBrida:2019mqg}
	{\scshape ALPHA} collaboration, M.~Dalla~Brida, R.~H\"ollwieser, F.~Knechtli,
	T.~Korzec, A.~Ramos and R.~Sommer, \emph{{Non-perturbative renormalization by
			decoupling}},
	\href{https://doi.org/10.1016/j.physletb.2020.135571}{\emph{Phys. Lett. B}
		{\bfseries 807} (2020) 135571}
	[\href{https://arxiv.org/abs/1912.06001}{{\ttfamily 1912.06001}}].
	
	\bibitem{DelDebbio:2021ryq}
	L.~Del~Debbio and A.~Ramos, \emph{{Lattice determinations of the strong
			coupling}},  \href{https://arxiv.org/abs/2101.04762}{{\ttfamily 2101.04762}}.
	
	\bibitem{Brambilla:2010pp}
	N.~Brambilla, X.~Garcia~i Tormo, J.~Soto and A.~Vairo, \emph{{Precision
			determination of $r_0\Lambda_{\overline{\mathrm{MS}}}$ from the QCD static
			energy}}, \href{https://doi.org/10.1103/PhysRevLett.105.212001}{\emph{Phys.
			Rev. Lett.} {\bfseries 105} (2010) 212001}
	[\href{https://arxiv.org/abs/1006.2066}{{\ttfamily 1006.2066}}].
	
	\bibitem{Asakawa:2015vta}
	M.~Asakawa, T.~Hatsuda, T.~Iritani, E.~Itou, M.~Kitazawa and H.~Suzuki,
	\emph{{Determination of Reference Scales for Wilson Gauge Action from
			Yang--Mills Gradient Flow}},
	\href{https://arxiv.org/abs/1503.06516}{{\ttfamily 1503.06516}}.
	
	\bibitem{Kitazawa:2016dsl}
	M.~Kitazawa, T.~Iritani, M.~Asakawa, T.~Hatsuda and H.~Suzuki, \emph{{Equation
			of State for SU(3) Gauge Theory via the Energy-Momentum Tensor under Gradient
			Flow}}, \href{https://doi.org/10.1103/PhysRevD.94.114512}{\emph{Phys. Rev. D}
		{\bfseries 94} (2016) 114512}
	[\href{https://arxiv.org/abs/1610.07810}{{\ttfamily 1610.07810}}].
	
	\bibitem{Ishikawa:2017xam}
	K.-I. Ishikawa, I.~Kanamori, Y.~Murakami, A.~Nakamura, M.~Okawa and R.~Ueno,
	\emph{{Non-perturbative determination of the $\Lambda$-parameter in the pure
			SU(3) gauge theory from the twisted gradient flow coupling}},
	\href{https://doi.org/10.1007/JHEP12(2017)067}{\emph{JHEP} {\bfseries 12}
		(2017) 067} [\href{https://arxiv.org/abs/1702.06289}{{\ttfamily
			1702.06289}}].
	
	\bibitem{Husung:2017qjz}
	N.~Husung, M.~Koren, P.~Krah and R.~Sommer, \emph{{SU(3) Yang Mills theory at
			small distances and fine lattices}},
	\href{https://doi.org/10.1051/epjconf/201817514024}{\emph{EPJ Web Conf.}
		{\bfseries 175} (2018) 14024}
	[\href{https://arxiv.org/abs/1711.01860}{{\ttfamily 1711.01860}}].
	
	\bibitem{DallaBrida:2019wur}
	M.~Dalla~Brida and A.~Ramos, \emph{{The gradient flow coupling at high-energy
			and the scale of SU(3) Yang\textendash{}Mills theory}},
	\href{https://doi.org/10.1140/epjc/s10052-019-7228-z}{\emph{Eur. Phys. J. C}
		{\bfseries 79} (2019) 720}
	[\href{https://arxiv.org/abs/1905.05147}{{\ttfamily 1905.05147}}].
	
	\bibitem{Nada:2020jay}
	A.~Nada and A.~Ramos, \emph{{An analysis of systematic effects in finite size
			scaling studies using the gradient flow}},
	\href{https://doi.org/10.1140/epjc/s10052-020-08759-1}{\emph{Eur. Phys. J. C}
		{\bfseries 81} (2021) 1} [\href{https://arxiv.org/abs/2007.12862}{{\ttfamily
			2007.12862}}].
	
	\bibitem{Husung:2020pxg}
	N.~Husung, A.~Nada and R.~Sommer, \emph{{Yang Mills short distance potential
			and perturbation theory}},
	\href{https://doi.org/10.22323/1.363.0263}{\emph{PoS} {\bfseries LATTICE2019}
		(2020) 263}.
	
	\bibitem{Bribian:2021cmg}
	E.~I. Bribian, J.~L.~D. Golan, M.~Garcia~Perez and A.~Ramos, \emph{{Memory
			efficient finite volume schemes with twisted boundary conditions}},
	\href{https://doi.org/10.1140/epjc/s10052-021-09718-0}{\emph{Eur. Phys. J. C}
		{\bfseries 81} (2021) 951}
	[\href{https://arxiv.org/abs/2107.03747}{{\ttfamily 2107.03747}}].
	
	\bibitem{Hasenfratz:2023bok}
	A.~Hasenfratz, C.~T. Peterson, J.~van Sickle and O.~Witzel,
	\emph{{\ensuremath{\Lambda} parameter of the SU(3) Yang-Mills theory from the
			continuous \ensuremath{\beta} function}},
	\href{https://doi.org/10.1103/PhysRevD.108.014502}{\emph{Phys. Rev. D}
		{\bfseries 108} (2023) 014502}
	[\href{https://arxiv.org/abs/2303.00704}{{\ttfamily 2303.00704}}].
	
	\bibitem{Narayanan:2006rf}
	R.~Narayanan and H.~Neuberger, \emph{{Infinite N phase transitions in continuum
			Wilson loop operators}},
	\href{https://doi.org/10.1088/1126-6708/2006/03/064}{\emph{JHEP} {\bfseries
			03} (2006) 064} [\href{https://arxiv.org/abs/hep-th/0601210}{{\ttfamily
			hep-th/0601210}}].
	
	\bibitem{Lohmayer:2011si}
	R.~Lohmayer and H.~Neuberger, \emph{{Continuous smearing of Wilson Loops}},
	\href{https://doi.org/10.22323/1.139.0249}{\emph{PoS} {\bfseries LATTICE2011}
		(2011) 249} [\href{https://arxiv.org/abs/1110.3522}{{\ttfamily 1110.3522}}].
	
	\bibitem{Luscher:2009eq}
	M.~Luscher, \emph{{Trivializing maps, the Wilson flow and the HMC algorithm}},
	\href{https://doi.org/10.1007/s00220-009-0953-7}{\emph{Commun. Math. Phys.}
		{\bfseries 293} (2010) 899}
	[\href{https://arxiv.org/abs/0907.5491}{{\ttfamily 0907.5491}}].
	
	\bibitem{Fritzsch:2013yxa}
	P.~Fritzsch, A.~Ramos and F.~Stollenwerk, \emph{{Critical slowing down and the
			gradient flow coupling in the Schr\"odinger functional}},
	\href{https://doi.org/10.22323/1.187.0461}{\emph{PoS} {\bfseries Lattice2013}
		(2014) 461} [\href{https://arxiv.org/abs/1311.7304}{{\ttfamily 1311.7304}}].
	
	\bibitem{Alles:1996vn}
	B.~Alles, G.~Boyd, M.~D'Elia, A.~Di~Giacomo and E.~Vicari, \emph{{Hybrid Monte
			Carlo and topological modes of full QCD}},
	\href{https://doi.org/10.1016/S0370-2693(96)01247-6}{\emph{Phys. Lett. B}
		{\bfseries 389} (1996) 107}
	[\href{https://arxiv.org/abs/hep-lat/9607049}{{\ttfamily hep-lat/9607049}}].
	
	\bibitem{DelDebbio:2004xh}
	L.~Del~Debbio, G.~M. Manca and E.~Vicari, \emph{{Critical slowing down of
			topological modes}},
	\href{https://doi.org/10.1016/j.physletb.2004.05.038}{\emph{Phys. Lett. B}
		{\bfseries 594} (2004) 315}
	[\href{https://arxiv.org/abs/hep-lat/0403001}{{\ttfamily hep-lat/0403001}}].
	
	\bibitem{Schaefer:2010hu}
	{\scshape ALPHA} collaboration, S.~Schaefer, R.~Sommer and F.~Virotta,
	\emph{{Critical slowing down and error analysis in lattice QCD simulations}},
	\href{https://doi.org/10.1016/j.nuclphysb.2010.11.020}{\emph{Nucl. Phys. B}
		{\bfseries 845} (2011) 93} [\href{https://arxiv.org/abs/1009.5228}{{\ttfamily
			1009.5228}}].
	
	\bibitem{Luscher:2014kea}
	M.~L\"uscher, \emph{{Step scaling and the Yang-Mills gradient flow}},
	\href{https://doi.org/10.1007/JHEP06(2014)105}{\emph{JHEP} {\bfseries 06}
		(2014) 105} [\href{https://arxiv.org/abs/1404.5930}{{\ttfamily 1404.5930}}].
	
	\bibitem{Albandea:2021lvl}
	D.~Albandea, P.~Hern\'andez, A.~Ramos and F.~Romero-L\'opez, \emph{{Topological
			sampling through windings}},
	\href{https://doi.org/10.1140/epjc/s10052-021-09677-6}{\emph{Eur. Phys. J. C}
		{\bfseries 81} (2021) 873}
	[\href{https://arxiv.org/abs/2106.14234}{{\ttfamily 2106.14234}}].
	
	\bibitem{Hasenbusch:2017unr}
	M.~Hasenbusch, \emph{{Fighting topological freezing in the two-dimensional
			$CP^{N-1}$ model}},
	\href{https://doi.org/10.1103/PhysRevD.96.054504}{\emph{Phys. Rev. D}
		{\bfseries 96} (2017) 054504}
	[\href{https://arxiv.org/abs/1706.04443}{{\ttfamily 1706.04443}}].
	
	\bibitem{Bonanno:2020hht}
	C.~Bonanno, C.~Bonati and M.~D'Elia, \emph{{Large-$N$ $SU(N)$ Yang-Mills
			theories with milder topological freezing}},
	\href{https://doi.org/10.1007/JHEP03(2021)111}{\emph{JHEP} {\bfseries 03}
		(2021) 111} [\href{https://arxiv.org/abs/2012.14000}{{\ttfamily
			2012.14000}}].
	
	\bibitem{Berni:2019bch}
	M.~Berni, C.~Bonanno and M.~D'Elia, \emph{{Large-$N$ expansion and
			$\theta$-dependence of $2d$ $CP^{N-1}$ models beyond the leading order}},
	\href{https://doi.org/10.1103/PhysRevD.100.114509}{\emph{Phys. Rev. D}
		{\bfseries 100} (2019) 114509}
	[\href{https://arxiv.org/abs/1911.03384}{{\ttfamily 1911.03384}}].
	
	\bibitem{Bonanno:2022yjr}
	C.~Bonanno, M.~D'Elia, B.~Lucini and D.~Vadacchino, \emph{{Towards glueball
			masses of large-N SU(N) pure-gauge theories without topological freezing}},
	\href{https://doi.org/10.1016/j.physletb.2022.137281}{\emph{Phys. Lett. B}
		{\bfseries 833} (2022) 137281}
	[\href{https://arxiv.org/abs/2205.06190}{{\ttfamily 2205.06190}}].
	
	\bibitem{Bonanno:2022hmz}
	C.~Bonanno, \emph{{Lattice determination of the topological susceptibility
			slope $\chi^\prime$ of $2d$ CP$^{N-1}$ models at large $N$}},
	\href{https://doi.org/10.1103/PhysRevD.107.014514}{\emph{Phys. Rev. D}
		{\bfseries 107} (2023) 014514}
	[\href{https://arxiv.org/abs/2212.02330}{{\ttfamily 2212.02330}}].
	
	\bibitem{Bonanno:2023hhp}
	C.~Bonanno, M.~D'Elia and L.~Verzichelli, \emph{{The
			\ensuremath{\theta}-dependence of the SU(N) critical temperature at large
			N}}, \href{https://doi.org/10.1007/JHEP02(2024)156}{\emph{JHEP} {\bfseries
			02} (2024) 156} [\href{https://arxiv.org/abs/2312.12202}{{\ttfamily
			2312.12202}}].
	
	\bibitem{Bonanno:2024ggk}
	C.~Bonanno, C.~Bonati, M.~Papace and D.~Vadacchino, \emph{{The
			$\theta$-dependence of the Yang-Mills spectrum from analytic continuation}},
	\href{https://doi.org/10.1007/JHEP05(2024)163}{\emph{JHEP} {\bfseries 05}
		(2024) 163} [\href{https://arxiv.org/abs/2402.03096}{{\ttfamily
			2402.03096}}].
	
	\bibitem{Bonanno:2024udh}
	C.~Bonanno, A.~Nada and D.~Vadacchino, \emph{{Mitigating topological freezing
			using out-of-equilibrium simulations}},
	\href{https://doi.org/10.1007/JHEP04(2024)126}{\emph{JHEP} {\bfseries 04}
		(2024) 126} [\href{https://arxiv.org/abs/2402.06561}{{\ttfamily
			2402.06561}}].
	
	\bibitem{Abbott:2024mix}
	R.~Abbott, D.~Boyda, D.~C. Hackett, G.~Kanwar, F.~Romero-L\'opez, P.~E.
	Shanahan et~al., \emph{{Practical applications of machine-learned flows on
			gauge fields}}, \href{https://doi.org/10.22323/1.453.0011}{\emph{PoS}
		{\bfseries LATTICE2023} (2024) 011}
	[\href{https://arxiv.org/abs/2404.11674}{{\ttfamily 2404.11674}}].
	
	\bibitem{Luscher:2011kk}
	M.~L{\"u}scher and S.~Schaefer, \emph{{Lattice QCD without topology barriers}},
	\href{https://doi.org/10.1007/JHEP07(2011)036}{\emph{JHEP} {\bfseries 07}
		(2011) 036} [\href{https://arxiv.org/abs/1105.4749}{{\ttfamily 1105.4749}}].
	
	\bibitem{Luscher:2012av}
	M.~L{\"u}scher and S.~Schaefer, \emph{{Lattice QCD with open boundary
			conditions and twisted-mass reweighting}},
	\href{https://doi.org/10.1016/j.cpc.2012.10.003}{\emph{Comput. Phys. Commun.}
		{\bfseries 184} (2013) 519}
	[\href{https://arxiv.org/abs/1206.2809}{{\ttfamily 1206.2809}}].
	
	\bibitem{DasilvaGolan:2023cjw}
	J.~L. Dasilva~Gol\'an, C.~Bonanno, M.~D'Elia, M.~Garc\'\i{}a~P\'erez and
	A.~Giorgieri, \emph{{The twisted gradient flow strong coupling with parallel
			tempering on boundary conditions}},
	\href{https://doi.org/10.22323/1.453.0354}{\emph{PoS} {\bfseries LATTICE2023}
		(2024) 354} [\href{https://arxiv.org/abs/2312.09212}{{\ttfamily
			2312.09212}}].
	
	\bibitem{Luscher:1991wu}
	M.~Luscher, P.~Weisz and U.~Wolff, \emph{{A Numerical method to compute the
			running coupling in asymptotically free theories}},
	\href{https://doi.org/10.1016/0550-3213(91)90298-C}{\emph{Nucl. Phys. B}
		{\bfseries 359} (1991) 221}.
	
	\bibitem{tHooft:1979rtg}
	G.~'t~Hooft, \emph{{A Property of Electric and Magnetic Flux in Nonabelian
			Gauge Theories}},
	\href{https://doi.org/10.1016/0550-3213(79)90595-9}{\emph{Nucl. Phys. B}
		{\bfseries 153} (1979) 141}.
	
	\bibitem{tHooft:1980kjq}
	G.~'t~Hooft, \emph{{Confinement and Topology in Nonabelian Gauge Theories}},
	{\emph{Acta Phys. Austriaca Suppl.} {\bfseries 22} (1980) 531}.
	
	\bibitem{Gonzalez-Arroyo:1982hyq}
	A.~Gonzalez-Arroyo and M.~Okawa, \emph{{The Twisted Eguchi-Kawai Model: A
			Reduced Model for Large N Lattice Gauge Theory}},
	\href{https://doi.org/10.1103/PhysRevD.27.2397}{\emph{Phys. Rev. D}
		{\bfseries 27} (1983) 2397}.
	
	\bibitem{Gonzalez-Arroyo:1982hwr}
	A.~Gonzalez-Arroyo and M.~Okawa, \emph{{A Twisted Model for Large $N$ Lattice
			Gauge Theory}},
	\href{https://doi.org/10.1016/0370-2693(83)90647-0}{\emph{Phys. Lett. B}
		{\bfseries 120} (1983) 174}.
	
	\bibitem{Gonzalez-Arroyo:2010omx}
	A.~Gonzalez-Arroyo and M.~Okawa, \emph{{Large $N$ reduction with the Twisted
			Eguchi-Kawai model}},
	\href{https://doi.org/10.1007/JHEP07(2010)043}{\emph{JHEP} {\bfseries 07}
		(2010) 043} [\href{https://arxiv.org/abs/1005.1981}{{\ttfamily 1005.1981}}].
	
	\bibitem{GarciaPerez:2014cmv}
	M.~Garcia~Perez, A.~Gonzalez-Arroyo and M.~Okawa, \emph{{Volume independence
			for Yang\textendash{}Mills fields on the twisted torus}},
	\href{https://doi.org/10.1142/S0217751X14450018}{\emph{Int. J. Mod. Phys. A}
		{\bfseries 29} (2014) 1445001}
	[\href{https://arxiv.org/abs/1406.5655}{{\ttfamily 1406.5655}}].
	
	\bibitem{GarciaPerez:2020gnf}
	M.~Garc\'\i{}a~P\'erez, \emph{{Prospects for large N gauge theories on the
			lattice}}, \href{https://doi.org/10.22323/1.363.0276}{\emph{PoS} {\bfseries
			LATTICE2019} (2020) 276} [\href{https://arxiv.org/abs/2001.10859}{{\ttfamily
			2001.10859}}].
	
	\bibitem{PhysRevLett.48.1063}
	T.~Eguchi and H.~Kawai, \emph{Reduction of dynamical degrees of freedom in the
		large-${N}$ gauge theory},
	\href{https://doi.org/10.1103/PhysRevLett.48.1063}{\emph{Phys. Rev. Lett.}
		{\bfseries 48} (1982) 1063}.
	
	\bibitem{Luscher:1982ma}
	M.~Lüscher, \emph{{Some Analytic Results Concerning the Mass Spectrum of
			Yang-Mills Gauge Theories on a Torus}},
	\href{https://doi.org/10.1016/0550-3213(83)90436-4}{\emph{Nucl. Phys. B}
		{\bfseries 219} (1983) 233}.
	
	\bibitem{Luscher:1992an}
	M.~Lüscher, R.~Narayanan, P.~Weisz and U.~Wolff, \emph{{The Schr\"odinger
			functional: A Renormalizable probe for nonAbelian gauge theories}},
	\href{https://doi.org/10.1016/0550-3213(92)90466-O}{\emph{Nucl. Phys. B}
		{\bfseries 384} (1992) 168}
	[\href{https://arxiv.org/abs/hep-lat/9207009}{{\ttfamily hep-lat/9207009}}].
	
	\bibitem{Ramos:2014kla}
	A.~Ramos, \emph{{The gradient flow running coupling with twisted boundary
			conditions}}, \href{https://doi.org/10.1007/JHEP11(2014)101}{\emph{JHEP}
		{\bfseries 11} (2014) 101} [\href{https://arxiv.org/abs/1409.1445}{{\ttfamily
			1409.1445}}].
	
	\bibitem{Bribian:2019ybc}
	E.~I. Bribian and M.~Garcia~Perez, \emph{{The twisted gradient flow coupling at
			one loop}}, \href{https://doi.org/10.1007/JHEP03(2019)200}{\emph{JHEP}
		{\bfseries 03} (2019) 200}
	[\href{https://arxiv.org/abs/1903.08029}{{\ttfamily 1903.08029}}].
	
	\bibitem{Creutz:1980zw}
	M.~Creutz, \emph{{Monte Carlo Study of Quantized $SU(2)$ Gauge Theory}},
	\href{https://doi.org/10.1103/PhysRevD.21.2308}{\emph{Phys. Rev. D}
		{\bfseries 21} (1980) 2308}.
	
	\bibitem{Kennedy:1985nu}
	A.~D. Kennedy and B.~J. Pendleton, \emph{{Improved Heat Bath Method for Monte
			Carlo Calculations in Lattice Gauge Theories}},
	\href{https://doi.org/10.1016/0370-2693(85)91632-6}{\emph{Phys. Lett. B}
		{\bfseries 156} (1985) 393}.
	
	\bibitem{Creutz:1987xi}
	M.~Creutz, \emph{{Overrelaxation and Monte Carlo Simulation}},
	\href{https://doi.org/10.1103/PhysRevD.36.515}{\emph{Phys. Rev. D} {\bfseries
			36} (1987) 515}.
	
	\bibitem{Luscher:2010iy}
	M.~L\"uscher, \emph{{Properties and uses of the Wilson flow in lattice QCD}},
	\href{https://doi.org/10.1007/JHEP08(2010)071}{\emph{JHEP} {\bfseries 08}
		(2010) 071} [\href{https://arxiv.org/abs/1006.4518}{{\ttfamily 1006.4518}}].
	
	\bibitem{Knechtli:2017xgy}
	{\scshape ALPHA} collaboration, F.~Knechtli, T.~Korzec, B.~Leder and G.~Moir,
	\emph{{Power corrections from decoupling of the charm quark}},
	\href{https://doi.org/10.1016/j.physletb.2017.10.025}{\emph{Phys. Lett. B}
		{\bfseries 774} (2017) 649}
	[\href{https://arxiv.org/abs/1706.04982}{{\ttfamily 1706.04982}}].
	
	\bibitem{Giusti:2018cmp}
	L.~Giusti and M.~L\"uscher, \emph{{Topological susceptibility at $T>T_{\rm c}$
			from master-field simulations of the SU(3) gauge theory}},
	\href{https://doi.org/10.1140/epjc/s10052-019-6706-7}{\emph{Eur. Phys. J. C}
		{\bfseries 79} (2019) 207}
	[\href{https://arxiv.org/abs/1812.02062}{{\ttfamily 1812.02062}}].
	
	\bibitem{Ce:2015qha}
	M.~C\`e, C.~Consonni, G.~P. Engel and L.~Giusti, \emph{{Non-Gaussianities in
			the topological charge distribution of the SU(3) Yang--Mills theory}},
	\href{https://doi.org/10.1103/PhysRevD.92.074502}{\emph{Phys. Rev. D}
		{\bfseries 92} (2015) 074502}
	[\href{https://arxiv.org/abs/1506.06052}{{\ttfamily 1506.06052}}].
	
	\bibitem{RevModPhys.53.43}
	D.~J. Gross, R.~D. Pisarski and L.~G. Yaffe, \emph{Qcd and instantons at finite
		temperature}, \href{https://doi.org/10.1103/RevModPhys.53.43}{\emph{Rev. Mod.
			Phys.} {\bfseries 53} (1981) 43}.
	
	\bibitem{Borsanyi:2015cka}
	S.~Bors{\'a}nyi, M.~Dierigl, Z.~Fodor, S.~D. Katz, S.~W. Mages,
	D.~N{\'o}gr{\'a}di et~al., \emph{{Axion cosmology, lattice QCD and the dilute
			instanton gas}},
	\href{https://doi.org/10.1016/j.physletb.2015.11.020}{\emph{Phys. Lett. B}
		{\bfseries 752} (2016) 175}
	[\href{https://arxiv.org/abs/1508.06917}{{\ttfamily 1508.06917}}].
	
	\bibitem{Petreczky:2016vrs}
	P.~Petreczky, H.-P. Schadler and S.~Sharma, \emph{{The topological
			susceptibility in finite temperature QCD and axion cosmology}},
	\href{https://doi.org/10.1016/j.physletb.2016.09.063}{\emph{Phys. Lett. B}
		{\bfseries 762} (2016) 498}
	[\href{https://arxiv.org/abs/1606.03145}{{\ttfamily 1606.03145}}].
	
	\bibitem{Borsanyi:2016ksw}
	S.~Bors{\'a}nyi, Z.~Fodor, J.~Guenther, K.-H. Kampert, S.~D. Katz, T.~Kawanai
	et~al., \emph{{Calculation of the axion mass based on high-temperature
			lattice quantum chromodynamics}},
	\href{https://doi.org/10.1038/nature20115}{\emph{Nature} {\bfseries 539}
		(2016) 69} [\href{https://arxiv.org/abs/1606.07494}{{\ttfamily 1606.07494}}].
	
	\bibitem{Jahn:2018dke}
	P.~T. Jahn, G.~D. Moore and D.~Robaina, \emph{{$\chi_{\textrm{top}}(T \gg
			T_{\textrm{c}})$ in pure-glue QCD through reweighting}},
	\href{https://doi.org/10.1103/PhysRevD.98.054512}{\emph{Phys. Rev. D}
		{\bfseries 98} (2018) 054512}
	[\href{https://arxiv.org/abs/1806.01162}{{\ttfamily 1806.01162}}].
	
	\bibitem{Bonati:2018blm}
	C.~Bonati, M.~D'Elia, G.~Martinelli, F.~Negro, F.~Sanfilippo and A.~Todaro,
	\emph{{Topology in full QCD at high temperature: a multicanonical approach}},
	\href{https://doi.org/10.1007/JHEP11(2018)170}{\emph{JHEP} {\bfseries 11}
		(2018) 170} [\href{https://arxiv.org/abs/1807.07954}{{\ttfamily
			1807.07954}}].
	
	\bibitem{Lombardo:2020bvn}
	M.~P. Lombardo and A.~Trunin, \emph{{Topology and axions in QCD}},
	\href{https://doi.org/10.1142/S0217751X20300100}{\emph{Int. J. Mod. Phys. A}
		{\bfseries 35} (2020) 2030010}
	[\href{https://arxiv.org/abs/2005.06547}{{\ttfamily 2005.06547}}].
	
	\bibitem{Borsanyi:2021gqg}
	S.~Bors{\'a}nyi and D.~Sexty, \emph{{Topological susceptibility of pure gauge
			theory using Density of States}},
	\href{https://doi.org/10.1016/j.physletb.2021.136148}{\emph{Phys. Lett. B}
		{\bfseries 815} (2021) 136148}
	[\href{https://arxiv.org/abs/2101.03383}{{\ttfamily 2101.03383}}].
	
	\bibitem{Athenodorou:2022aay}
	A.~Athenodorou, C.~Bonanno, C.~Bonati, G.~Clemente, F.~D'Angelo, M.~D'Elia
	et~al., \emph{{Topological susceptibility of N$_{f}$ = 2 + 1 QCD from
			staggered fermions spectral projectors at high temperatures}},
	\href{https://doi.org/10.1007/JHEP10(2022)197}{\emph{JHEP} {\bfseries 10}
		(2022) 197} [\href{https://arxiv.org/abs/2208.08921}{{\ttfamily
			2208.08921}}].
	
	\bibitem{Bonanno:2022dru}
	C.~Bonanno, M.~D'Elia and F.~Margari, \emph{{Topological susceptibility of the
			2D CP$^1$ or O(3) nonlinear \ensuremath{\sigma} model: Is it divergent or
			not?}}, \href{https://doi.org/10.1103/PhysRevD.107.014515}{\emph{Phys. Rev.
			D} {\bfseries 107} (2023) 014515}
	[\href{https://arxiv.org/abs/2208.00185}{{\ttfamily 2208.00185}}].
	
	\bibitem{Lucini:2023irm}
	B.~Lucini, D.~Mason, M.~Piai, E.~Rinaldi and D.~Vadacchino, \emph{{First-order
			phase transitions in Yang-Mills theories and the density of state method}},
	\href{https://doi.org/10.1103/PhysRevD.108.074517}{\emph{Phys. Rev. D}
		{\bfseries 108} (2023) 074517}
	[\href{https://arxiv.org/abs/2305.07463}{{\ttfamily 2305.07463}}].
	
	\bibitem{Bonati:2015sqt}
	C.~Bonati, M.~D'Elia and A.~Scapellato, \emph{{$\theta$ dependence in $SU(3)$
			Yang-Mills theory from analytic continuation}},
	\href{https://doi.org/10.1103/PhysRevD.93.025028}{\emph{Phys. Rev. D}
		{\bfseries 93} (2016) 025028}
	[\href{https://arxiv.org/abs/1512.01544}{{\ttfamily 1512.01544}}].
	
	\bibitem{Witten:1979vv}
	E.~Witten, \emph{{Current Algebra Theorems for the $U(1)$ Goldstone Boson}},
	\href{https://doi.org/10.1016/0550-3213(79)90031-2}{\emph{Nucl. Phys. B}
		{\bfseries 156} (1979) 269}.
	
	\bibitem{Veneziano:1979ec}
	G.~Veneziano, \emph{{$U(1)$ Without Instantons}},
	\href{https://doi.org/10.1016/0550-3213(79)90332-8}{\emph{Nucl. Phys. B}
		{\bfseries 159} (1979) 213}.
	
	\bibitem{Alles:1996nm}
	B.~Alles, M.~D'Elia and A.~Di~Giacomo, \emph{{Topological susceptibility at
			zero and finite T in SU(3) Yang-Mills theory}},
	\href{https://doi.org/10.1016/S0550-3213(97)00205-8}{\emph{Nucl. Phys. B}
		{\bfseries 494} (1997) 281}
	[\href{https://arxiv.org/abs/hep-lat/9605013}{{\ttfamily hep-lat/9605013}}].
	
	\bibitem{Alles:1997qe}
	B.~Alles, M.~D'Elia and A.~Di~Giacomo, \emph{{Topology at zero and finite $T$
			in $SU(2)$ Yang--Mills theory}},
	\href{https://doi.org/10.1016/S0370-2693(97)01059-9}{\emph{Phys. Lett. B}
		{\bfseries 412} (1997) 119}
	[\href{https://arxiv.org/abs/hep-lat/9706016}{{\ttfamily hep-lat/9706016}}].
	
	\bibitem{DelDebbio:2002xa}
	L.~Del~Debbio, H.~Panagopoulos and E.~Vicari, \emph{{$\theta$ dependence of
			$SU(N)$ gauge theories}},
	\href{https://doi.org/10.1088/1126-6708/2002/08/044}{\emph{JHEP} {\bfseries
			08} (2002) 044} [\href{https://arxiv.org/abs/hep-th/0204125}{{\ttfamily
			hep-th/0204125}}].
	
	\bibitem{DelDebbio:2004ns}
	L.~Del~Debbio, L.~Giusti and C.~Pica, \emph{{Topological susceptibility in the
			$SU(3)$ gauge theory}},
	\href{https://doi.org/10.1103/PhysRevLett.94.032003}{\emph{Phys. Rev. Lett.}
		{\bfseries 94} (2005) 032003}
	[\href{https://arxiv.org/abs/hep-th/0407052}{{\ttfamily hep-th/0407052}}].
	
	\bibitem{DElia:2003zne}
	M.~D'Elia, \emph{{Field theoretical approach to the study of theta dependence
			in Yang-Mills theories on the lattice}},
	\href{https://doi.org/10.1016/S0550-3213(03)00311-0}{\emph{Nucl. Phys. B}
		{\bfseries 661} (2003) 139}
	[\href{https://arxiv.org/abs/hep-lat/0302007}{{\ttfamily hep-lat/0302007}}].
	
	\bibitem{Lucini:2004yh}
	B.~Lucini, M.~Teper and U.~Wenger, \emph{{Topology of $SU(N)$ gauge theories at
			$T \simeq 0$ and $T \simeq T_c$}},
	\href{https://doi.org/10.1016/j.nuclphysb.2005.02.037}{\emph{Nucl. Phys. B}
		{\bfseries 715} (2005) 461}
	[\href{https://arxiv.org/abs/hep-lat/0401028}{{\ttfamily hep-lat/0401028}}].
	
	\bibitem{Giusti:2007tu}
	L.~Giusti, S.~Petrarca and B.~Taglienti, \emph{{$\theta$ dependence of the
			vacuum energy in the $SU(3)$ gauge theory from the lattice}},
	\href{https://doi.org/10.1103/PhysRevD.76.094510}{\emph{Phys. Rev. D}
		{\bfseries 76} (2007) 094510}
	[\href{https://arxiv.org/abs/0705.2352}{{\ttfamily 0705.2352}}].
	
	\bibitem{Vicari:2008jw}
	E.~Vicari and H.~Panagopoulos, \emph{{$\theta$ dependence of $SU(N)$ gauge
			theories in the presence of a topological term}},
	\href{https://doi.org/10.1016/j.physrep.2008.10.001}{\emph{Phys. Rept.}
		{\bfseries 470} (2009) 93} [\href{https://arxiv.org/abs/0803.1593}{{\ttfamily
			0803.1593}}].
	
	\bibitem{Panagopoulos:2011rb}
	H.~Panagopoulos and E.~Vicari, \emph{{The $4D$ $SU(3)$ gauge theory with an
			imaginary $\theta$ term}},
	\href{https://doi.org/10.1007/JHEP11(2011)119}{\emph{JHEP} {\bfseries 11}
		(2011) 119} [\href{https://arxiv.org/abs/1109.6815}{{\ttfamily 1109.6815}}].
	
	\bibitem{Bonati:2013tt}
	C.~Bonati, M.~D'Elia, H.~Panagopoulos and E.~Vicari, \emph{{Change of
			\ensuremath{\theta} Dependence in $4D$ $SU(N)$ Gauge Theories Across the
			Deconfinement Transition}},
	\href{https://doi.org/10.1103/PhysRevLett.110.252003}{\emph{Phys. Rev. Lett.}
		{\bfseries 110} (2013) 252003}
	[\href{https://arxiv.org/abs/1301.7640}{{\ttfamily 1301.7640}}].
	
	\bibitem{Ce:2016awn}
	M.~C\`e, M.~Garcia~Vera, L.~Giusti and S.~Schaefer, \emph{{The topological
			susceptibility in the large-$N$ limit of SU($N$) Yang-Mills theory}},
	\href{https://doi.org/10.1016/j.physletb.2016.09.029}{\emph{Phys. Lett. B}
		{\bfseries 762} (2016) 232}
	[\href{https://arxiv.org/abs/1607.05939}{{\ttfamily 1607.05939}}].
	
	\bibitem{Berkowitz:2015aua}
	E.~Berkowitz, M.~I. Buchoff and E.~Rinaldi, \emph{{Lattice QCD input for axion
			cosmology}}, \href{https://doi.org/10.1103/PhysRevD.92.034507}{\emph{Phys.
			Rev. D} {\bfseries 92} (2015) 034507}
	[\href{https://arxiv.org/abs/1505.07455}{{\ttfamily 1505.07455}}].
	
	\bibitem{Bonati:2016tvi}
	C.~Bonati, M.~D'Elia, P.~Rossi and E.~Vicari, \emph{{$\theta$ dependence of 4D
			$SU(N)$ gauge theories in the large-$N$ limit}},
	\href{https://doi.org/10.1103/PhysRevD.94.085017}{\emph{Phys. Rev. D}
		{\bfseries 94} (2016) 085017}
	[\href{https://arxiv.org/abs/1607.06360}{{\ttfamily 1607.06360}}].
	
	\bibitem{Bonati:2018rfg}
	C.~Bonati, M.~Cardinali and M.~D'Elia, \emph{{$\theta$ dependence in trace
			deformed $SU(3)$ Yang-Mills theory: a lattice study}},
	\href{https://doi.org/10.1103/PhysRevD.98.054508}{\emph{Phys. Rev. D}
		{\bfseries 98} (2018) 054508}
	[\href{https://arxiv.org/abs/1807.06558}{{\ttfamily 1807.06558}}].
	
	\bibitem{Bonati:2019kmf}
	C.~Bonati, M.~Cardinali, M.~D'Elia and F.~Mazziotti, \emph{{$\theta$-dependence
			and center symmetry in Yang-Mills theories}},
	\href{https://doi.org/10.1103/PhysRevD.101.034508}{\emph{Phys. Rev. D}
		{\bfseries 101} (2020) 034508}
	[\href{https://arxiv.org/abs/1912.02662}{{\ttfamily 1912.02662}}].
	
	\bibitem{Gonzalez-Arroyo:1995ynx}
	A.~Gonzalez-Arroyo and P.~Martinez, \emph{{Investigating Yang-Mills theory and
			confinement as a function of the spatial volume}},
	\href{https://doi.org/10.1016/0550-3213(95)00601-X}{\emph{Nucl. Phys. B}
		{\bfseries 459} (1996) 337}
	[\href{https://arxiv.org/abs/hep-lat/9507001}{{\ttfamily hep-lat/9507001}}].
	
	\bibitem{Gonzalez-Arroyo:2023kqv}
	A.~Gonzalez-Arroyo, \emph{{On the fractional instanton liquid picture of the
			Yang-Mills vacuum and Confinement}},
	\href{https://arxiv.org/abs/2302.12356}{{\ttfamily 2302.12356}}.
	
	\bibitem{Gross:1980br}
	D.~J. Gross, R.~D. Pisarski and L.~G. Yaffe, \emph{{QCD and Instantons at
			Finite Temperature}},
	\href{https://doi.org/10.1103/RevModPhys.53.43}{\emph{Rev. Mod. Phys.}
		{\bfseries 53} (1981) 43}.
	
	\bibitem{Schafer:1996wv}
	T.~Sch\"afer and E.~V. Shuryak, \emph{{Instantons in QCD}},
	\href{https://doi.org/10.1103/RevModPhys.70.323}{\emph{Rev. Mod. Phys.}
		{\bfseries 70} (1998) 323}
	[\href{https://arxiv.org/abs/hep-ph/9610451}{{\ttfamily hep-ph/9610451}}].
	
	\bibitem{vanBaal:1984ra}
	P.~van Baal, \emph{{Twisted Boundary Conditions: A Nonperturbative Probe for
			Pure Nonabelian Gauge Theories}},  {P}h.{D}. {T}hesis, Utrecht U., 7, 1984.
	
	\bibitem{RTN:1993ilw}
	{\scshape RTN} collaboration, M.~Garcia~Perez et~al., \emph{{Instanton like
			contributions to the dynamics of Yang-Mills fields on the twisted torus}},
	\href{https://doi.org/10.1016/0370-2693(93)91069-Y}{\emph{Phys. Lett. B}
		{\bfseries 305} (1993) 366}
	[\href{https://arxiv.org/abs/hep-lat/9302007}{{\ttfamily hep-lat/9302007}}].
	
	\bibitem{GarciaPerez:1993jw}
	M.~Garcia~Perez, A.~Gonzalez-Arroyo and P.~Martinez, \emph{{From perturbation
			theory to confinement: How the string tension is built up}},
	\href{https://doi.org/10.1016/0920-5632(94)90352-2}{\emph{Nucl. Phys. B Proc.
			Suppl.} {\bfseries 34} (1994) 228}
	[\href{https://arxiv.org/abs/hep-lat/9312066}{{\ttfamily hep-lat/9312066}}].
	
	\bibitem{vanBaal:2000zc}
	P.~van Baal, \emph{{QCD in a finite volume}},
	\href{https://arxiv.org/abs/hep-ph/0008206}{{\ttfamily hep-ph/0008206}}.
	
	\bibitem{Unsal:2020yeh}
	M.~\"Unsal, \emph{{Strongly coupled QFT dynamics via TQFT coupling}},
	\href{https://doi.org/10.1007/JHEP11(2021)134}{\emph{JHEP} {\bfseries 11}
		(2021) 134} [\href{https://arxiv.org/abs/2007.03880}{{\ttfamily
			2007.03880}}].
	
	\bibitem{Cox:2021vsa}
	A.~A. Cox, E.~Poppitz and F.~D. Wandler, \emph{{The mixed 0-form/1-form anomaly
			in Hilbert space: pouring the new wine into old bottles}},
	\href{https://doi.org/10.1007/JHEP10(2021)069}{\emph{JHEP} {\bfseries 10}
		(2021) 069} [\href{https://arxiv.org/abs/2106.11442}{{\ttfamily
			2106.11442}}].
	
	\bibitem{Nair:2022yqi}
	V.~P. Nair and R.~D. Pisarski, \emph{{Fractional topological charge in SU(N)
			gauge theories without dynamical quarks}},
	\href{https://doi.org/10.1103/PhysRevD.108.074007}{\emph{Phys. Rev. D}
		{\bfseries 108} (2023) 074007}
	[\href{https://arxiv.org/abs/2206.11284}{{\ttfamily 2206.11284}}].
	
\end{thebibliography}
\end{document}